\documentclass[fleqn,usenatbib]{mnras}

\usepackage{newtxtext,newtxmath}

\usepackage[T1]{fontenc}

\DeclareRobustCommand{\VAN}[3]{#2}
\let\VANthebibliography\thebibliography
\def\thebibliography{\DeclareRobustCommand{\VAN}[3]{##3}\VANthebibliography}

\usepackage{graphicx}	
\usepackage{amsmath}	
\usepackage{placeins}   %
\usepackage{multirow}

\title[Neutrino mass constraints with peculiar velocities]{Using peculiar velocity surveys to constrain neutrino masses}

\author[A. M. Whitford et al.]{
Abb\'e M. Whitford,$^{1}$\thanks{E-mail: abbe.whitford@gmail.com}
Cullan Howlett,$^{1}$
Tamara M. Davis$^{1}$
\\
$^{1}$School of Mathematics and Physics, University of Queensland, Brisbane QLD 4072, Australia \\
}

\date{Accepted XXX. Received YYY; in original form ZZZ}

\pubyear{2021}

\begin{document}
\label{firstpage}
\pagerange{\pageref{firstpage}--\pageref{lastpage}}
\maketitle

\begin{abstract}
The presence of massive neutrinos in the early Universe is expected to have influenced the observed distribution of galaxies, and their observed motions. In this work, we explore whether measurements of galaxy peculiar velocities could allow us to improve upon neutrino mass constraints from galaxy redshift surveys alone. Using Fisher matrix forecasts, we show that the galaxy peculiar motions \textit{do} contain information on the sum of the masses of neutrinos $\sum m_{\nu}$, and that this information can be used to improve upon constraints that may be obtained from low-redshift galaxy surveys ($z<0.5$) combined with \textit{Planck} measurements of the Cosmic Microwave Background. Compared to the full constraining power offered by \textit{Planck} and higher redshift DESI data, we find that the benefit of including peculiar velocities only marginally improves neutrino mass constraints. However, when one does not include information from \textit{Planck}, our results show that the inclusion of peculiar velocity measurements can substantially improve upon the constraints from redshift surveys alone, and that in some cases the addition of further data from high density peculiar velocity surveys is more successful at overcoming the sample variance than adding further data from redshifts only. We demonstrate that it may be possible to achieve upper bounds of $\sum m_{\nu} \approx $0.14 eV (68\% confidence limit), from upcoming spectroscopic galaxy surveys alone, as long as the peculiar velocity data is available, an improvement of $\sim$14\% over redshift surveys alone, but not as strong as when CMB data is included. 

\end{abstract}


\section{Introduction }
\label{intro}

Neutrinos are some of the most mysterious particles in the standard model of particle physics. As they only interact via the weak nuclear force and gravity, they are notoriously difficult to detect.  While they were initially believed to be massless particles, as is predicted by the standard model, the discovery of neutrino oscillations in 1998 \citep{fukuda1998evidence} and subsequent detections in other reactor, particle accelerator, solar and atmospheric experiments (e.g., \citealt{ikeda2008solar, abe2018atmospheric, acero2019first, adey2018measurement}) has demonstrated that at least two of the three standard model neutrinos must be massive. However, these masses are extremely small -- the current upper bounds on the sum of masses across all three neutrino species is constrained to be at least half a million times smaller than the rest mass of the electron \citep{aghanim2020planck, gerbino2018status}. 
\par 
While various oscillation experiments have allowed for constraints on the squared differences in mass (i.e. mass splittings) between each neutrino mass eigenstate, which puts a lower bound on the value of the sum of masses, $\sum m_{\nu}$ (depending on how the mass splittings are ordered, \citealt{esteban2020fate}), no experiment yet exists that can put a bound of the mass of any individual eigenstate. However, cosmological probes have been able to put increasingly tight upper bounds on $\sum m_{\nu}$; recent measurements of the Cosmic Microwave Background (CMB) by the \textit{Planck} satellite, in combination with measurements of baryon acoustic oscillations (BAOs) and Lyman-$\alpha$ forest data give $\sum m_{\nu} < 0.12$ eV \citep{palanque2015neutrino,aghanim2020planck}. A slightly tighter bound comes from \textit{Planck} 2018 data combined with data from BAOs, Supernovae Ia (SNe Ia), Redshift space distortions, galaxy weak lensing and clustering data, $\sum m_{\nu} < 0.111$ eV \citep{alam2021completed}. A combined analysis of cosmological information and neutrino oscillation experiments hence allows for constraints on the individual masses of the three standard model neutrinos, \citep[e.g.][]{Stoecker2021}, although as of yet there has been no actual measurement of either the $\sum m_{\nu}$ or a non-zero lightest neutrino mass.
\par 
The ability of cosmological probes, in particular the CMB power spectrum and the matter power spectrum, to constrain neutrino masses, is mainly due to two effects on these probes induced by massive neutrinos. Neutrinos are relativistic almost immediately after the Big Bang, due to the high temperature thermal plasma they interacted with, and thus they behaved like a species of radiation. However, neutrinos with mass should eventually transition to non-relativistic due to the Universe expanding and cooling. Hence, the presence of massive neutrinos alters the expansion rate of the Universe as they contribute to the total radiation density $\Omega_{\text{r}}$ at early times and total matter energy density $\Omega_{\text{m}}$ at later times. Firstly, this has the effect of changing the epoch of matter-radiation (MR) equality \citep{lesgourgues2012neutrino},
\begin{equation}
    \frac{a_{\text{eq}}}{a_0} = \frac{\Omega_{\text{r}}}{\Omega_{\text{m}}},
\end{equation}
where $a_{0}$ is the scale factor at the current day and $a_{\text{eq}}$ is the scale factor at the epoch of MR equality; if they are still relativistic at MR equality, they add to the radiation density, $\Omega_{\text{r}} = \Omega_{\gamma} + \Omega_{\nu}$, where $\Omega_{\gamma}$ is the energy density of photons and $\Omega_{\nu}$ is the energy density of neutrinos. Otherwise, $\Omega_{\nu}$ contributes to $\Omega_{\text{m}}$ at MR equality and afterwards (all energy densities $\Omega$ are normalized by the critical density of the Universe, $\rho_{\text{cr}}$). This in turn changes the position of peaks in the primary CMB anisotropies by altering the angular diameter distance to the CMB, and also the position of the BAO peaks and the turnover in the matter power spectrum; a larger value of $\Omega_{\nu}$, that contributes to $\Omega_{\text{r}}$ at the epoch of MR equality, tends to overall push the CMB anisotropies in the power spectrum to smaller values of $\ell$, and the same occurs for the BAO peaks and turnover in the matter power spectrum which are pushed to smaller $k$ modes. Their contribution to the $\Omega_{\text{m}}$ after the epoch of recombination is also detectable in the secondary CMB anisotropies that are influenced by the expansion rate of the Universe at different times. \par
\begin{figure}
    \centering
    \includegraphics[scale=0.475]{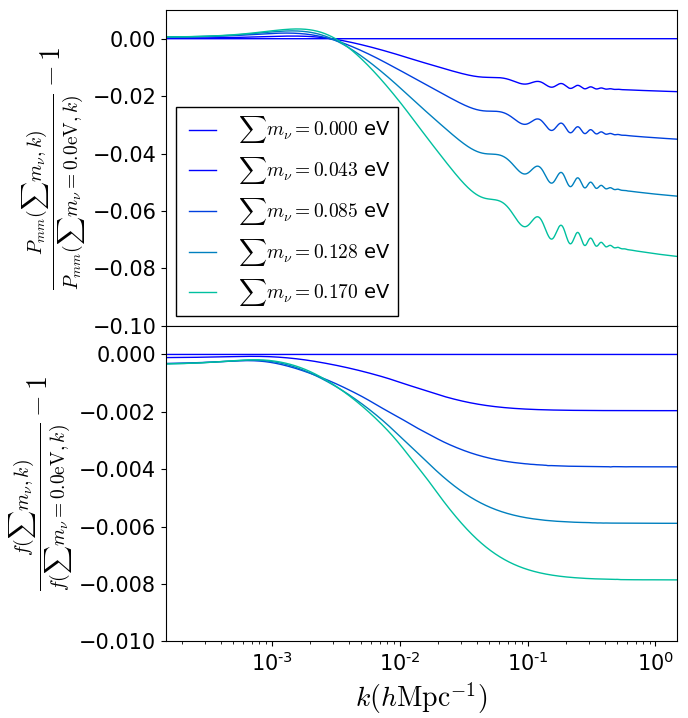}
    \caption{Top panel: The fractional difference in the $z=0$ linear matter power spectrum $P(k)$ with varying $\sum m_{\nu}$ compared to a matter power spectrum with $\sum m_{\nu} = 0.0$ eV. Lower panel: the same, except for the growth rate of structure, $f(k)$. Other cosmological parameters remain constant except $\Omega_{\text{b}}$ and $\Omega_{\text{cdm}}$, the energy density of baryons and cold dark matter (cdm) respectively, which maintain a constant ratio such that $\Omega_{\text{m}} = \Omega_{\text{b}} + \Omega_{\text{cdm}} + \Omega_{\nu}$ also remains constant. This matter power spectra were generated in this plot with \texttt{CLASS} \citep{blas2011cosmic}.}
    \label{figure1_powerspectrum}
\end{figure}

 
Secondly, due to their weak interactions, neutrinos decoupled from the baryon-photon fluid almost immediately after the Big Bang and began to fly freely through the Universe \citep{lesgourgues2012neutrino, riemer2013half}. The effect of this is that neutrinos are expected to have influenced the distribution of matter, since they do not cluster, and thus they inhibit the growth of structures on scales smaller than their free-streaming length by reducing the gravitational potential. As neutrinos transition to being non-relativistic, their free-streaming length, which is the typical distance over which neutrinos may travel, starts to overall decrease with expansion of the Universe as their thermal velocity decays faster than the free-streaming length can increase due to the expansion of the Universe. This free-streaming length passes through a minimum wavenumber, which creates a characteristic scale in the matter power spectrum, given by
\begin{equation}
    k_{\text{nr}} \approx 0.018 \sqrt{\Omega_{\text{m}}}\biggl(\frac{m_{\nu}}{1\mathrm{eV}}\biggl)^{1/2}h\,\text{Mpc}^{-1},
\end{equation}
\citep{lesgourgues2014neutrino}. At length scales below this, structure growth is suppressed. This in turn has an impact on the CMB anisotropies and the matter power spectrum. Although, strictly speaking, the time of the non-relativistic transition and the free-streaming length scale is different for each massive neutrino specie, in practice only the cumulative effect of all neutrino species is measurable \citep{Archidiacono2020}. The effects of altering $\sum m_{\nu}$ on the matter power spectrum is demonstrated in the top panel of Figure \ref{figure1_powerspectrum}.
\par 
Upcoming projects aim to obtain the first true measurement (as opposed to upper limits) on $\sum m_{\nu}$, including the Dark Energy Spectroscopic Instrument (DESI), the European Space Agency Cosmic Vision mission Euclid (Euclid) and Stage-4 CMB experiments (CMB-S4) \citep{laureijs2012euclid, aghamousa2016desi, abazajian2019cmb}. Various forecasts for constraints on $\sum m_{\nu}$ by some of these projects predict high precision 1--$\sigma$ uncertainties on $\sum m_{\nu}$, that are as low as $\sigma_{\sum m_{\nu}} = 0.01-0.025$ eV \citep{allison2015towards,aghamousa2016desi, sartoris2016next,sprenger2019cosmology}, when future data from these projects is combined or used to augment current data.
\par 
Current bounds on $\sum m_{\nu}$ and the above forecasts for $\sum m_{\nu}$ in the literature focus on using information from the CMB, or information from the matter power spectrum. The latter is accessible in several ways, including via the  large scale distribution of galaxies and redshift space distortions (RSDs), measurements of the Lyman-$\alpha$ forest, or weak gravitational lensing of galaxies. Other upcoming projects, such as the Square Kilometre Array (SKA),\footnote{https://www.skatelescope.org/science/} will also measure the matter power spectrum with the distribution of hydrogen in the Universe, obtained by 21 cm intensity mapping. However, all of these previous studies have considered only the \textit{spatial distribution} of galaxies, while the free-streaming effects of neutrinos can also reasonably be expected to have influenced the present day \textit{motions} of galaxies, i.e. galaxy peculiar velocities. These have not previously been considered as a probe for constraining $\sum m_{\nu}$. We expect measurements of the peculiar motions of galaxies to also contain information on $\sum m_{\nu}$, since these velocities are sourced by the relative gravitational potential between different regions of the Universe, including potentials that have been altered on scales less than the neutrino free streaming length scale, as described previously. Furthermore, various works have shown that bounds on cosmological parameters can be improved relative to those obtained by redshift surveys alone, when redshift survey data is used in combination with peculiar velocity data in the same cosmological volume \citep{koda2014peculiar, howlett2017cosmological, howlett2017measuring, quartin20216x2pt, amendola2021measuring, castro2016turning, agrawal2019constraining, Burkey2004}.
\par 
Hence, this work aims to answer two questions for the first time:
\begin{enumerate}
    \item {Do realistic measurements of the peculiar motions of galaxies improve constraints on $\sum m_{\nu}$ compared to redshift measurements alone?} 
    \item {Is the improvement (if any) we may obtain on $\sum m_{\nu}$ from peculiar velocity (PV) data actually useful given the tight constraints we can already obtain from CMB and galaxy survey data over a wide range of redshifts.}
\end{enumerate}
In order to address these questions, this paper is laid out as follows. In Section~\ref{sec:PV}, we briefly introduce peculiar velocities before discussing in more depth the motivation for using peculiar velocity surveys to try obtain improved constraints on $\sum m_{\nu}$. Our Fisher matrix forecasting method for attempting to answer the above two questions is explained in Section~\ref{sec:fisher}. In Section~\ref{sec:surveys}, we discuss surveys that will measure peculiar velocities that may allow for us to derive tight constraints on $\sum m_{\nu}$. In Section~\ref{sec:results} we present our forecasts for these surveys. We discuss the effects of nuisance parameters and systematics on our results in Section~\ref{sec:discussion} and conclude in Section~\ref{sec:conclusions}.

The code to produce the forecasts shown in this paper is available at \href{https://github.com/abbew25/PV_galaxy_survey_forecasts_neutrinos}{https://github.com/abbew25/PV\_galaxy\_survey\_forecasts\_neutrinos}.

\section{Peculiar velocity surveys}
\label{sec:PV}

\subsection{Overview}

All galaxies can be thought of having two components of motion. These components are the galaxy's motion due to the expansion of space, and their motion due to their interactions with local gravitational fields, called peculiar motions or peculiar velocities. A galaxy's redshift, obtained from photometry or spectroscopy, is related to the total velocity of the galaxy in the radial direction, which can be written as
\begin{equation}
    v_{\text{total}} = v_{\text{pec}} + v_{\text{rec}},
\end{equation}
where $v_{\text{pec}}$ is the peculiar velocity, and $v_{\text{rec}}$ is the recessional velocity due to the expansion of space. From an observed redshift alone, it is not possible to isolate these two components of motion (except in a statistical sense when averaging over many thousands of galaxies, i.e., using RSDs), since a galaxy's observed redshift $z_{\text{obs}}$ is related to the cosmological redshift $z_{\text{rec}}$ and the peculiar redshift $z_{\text{pec}}$ by
\begin{equation}
    1 + z_{\text{obs}} = (1 + z_{\text{pec}})(1 + z_{\text{rec}}).
\end{equation}
However, if one can obtain a model independent measurement of the distance to a galaxy, for example, using the Fundamental Plane \citep{djorgovski1987fundamental}, the Tully-Fisher relation \citep{tully1977new} or Type Ia SNe \citep{phillips1993absolute}, this can be related to $z_{\text{rec}}$ by assuming a cosmological model, thus isolating $z_{\text{pec}}$ from $z_{\text{obs}}$. The knowledge of $z_{\text{pec}}$ allows one to determine the peculiar motion of the galaxy in the radial direction.

\subsection{Motivating combining PV surveys with redshift surveys}

Surveys that obtain observed redshifts for galaxies, which we will consider from here on as equivalently obtaining information about the galaxy density field, $\delta_g(k)$, allow us to measure the power spectrum of galaxies, $P_{gg}(k)$. This is related to the matter power spectrum, $P_{mm}(k)$, to a first order approximation as
\begin{equation}
    P_{gg}(k) = b^2_gP_{mm}(k),
\end{equation}
where $b_g$ is the linear bias factor. The bias factor accounts for the fact that galaxies are more likely to form in regions where the density of dark matter is the greatest and hence do not probe the underlying distribution of matter uniformly. In contrast, the galaxy line-of-sight velocity field $u(k)$, which can be probed with the galaxy velocity power spectrum $P_{uu}(k)$ constructed from a peculiar velocity survey, is believed to be unbiased (at least on linear scales) compared to the underlying galaxy velocity divergence power spectrum, $P_{\theta \theta}(k)$ \citep{Desjacques2010,zheng2015determination}. The velocity divergence field $\theta(k)$ is defined in linear theory such that $\theta(k) \equiv \delta(k)$, where $\delta(k)$ is the underlying matter field, and thus the velocity power spectrum is an unbiased probe of the underlying matter field on large scales. Thus we expect that measurements of both the galaxy density field from redshift surveys, in addition to measurements of the velocity field from peculiar velocity (PV) surveys can help to break the degeneracy between cosmological parameters and galaxy bias that afflicts redshift surveys.
\par
When we take into account the effects of redshift space distortions (RSDs), our measured redshift space galaxy power spectrum and galaxy velocity power spectrum become functions of both scale and the angle between the line-of-sight and galaxy separation vector $\mu$. These power spectra can be modelled as \citep{koda2014peculiar}
\begin{align}
    P_{gg}(k) & = (b^2_g + f^2(k)\mu^2)^2 D^2_gP_{mm}(k), \\
    P_{uu}(k) & = \left( \frac{aH(a)\mu f(k)}{k} \right)^2 D^2_u P_{\theta \theta}(k),
\end{align}
where $H(a)$ is the Hubble parameter, $f(k)$ is the linear growth rate of structure and $D_g$ and $D_u$ are damping terms for each power spectrum respectively to account for the non-linear motions of galaxies within their host clusters (the so called Fingers-of-God (FoG) effect) \citep{koda2014peculiar}. 

The linear growth rate of structure, $f(k)$, is a measure of the rate of growth of a density perturbation and is related directly to the speed at which galaxies fall into clusters. This is often assumed to be constant with respect to scale $k$, $f \approx \Omega_{\text{m}}^{\gamma}(z)$, where $\gamma \approx 0.55$ in General Relativity. However, the presence of massive neutrinos in the early Universe introduces a scale dependence to the growth of structure, which we incorporate in our analysis. The effect of changing $\sum m_{\nu}$ on $f(k)$ is shown in Figure \ref{figure1_powerspectrum}; how this can be computed numerically is discussed in Section~\ref{sec:numericalimplementation}.

The factor of $(b_g + f\mu^2)$ in the equation for $P_{gg}(k)$ is called the Kaiser factor and accounts for the statistical `squashing' along the line-of-sight of the power spectrum from galaxies falling towards dense regions of the Universe \citep{kaiser1987clustering}. The factor of $\frac{aH(a)\mu f(k)}{k}$ in the equation for $P_{uu}(k)$ arises due to the definition of the linear velocity divergence field, $\theta(k)$ and as we are only able to measure peculiar velocities along the line-of-sight. $P_{gg}(k)$ and $P_{uu}(k)$ also each have a different dependence on $\mu$, which helps to constrain cosmological parameters. The cross-correlation of the density and velocity fields can be written as \citep{koda2014peculiar}
\begin{align}
    P_{ug}(k) & = P_{gu}(k) \\ \nonumber
    & =  \left( \frac{aH(a)\mu f(k)}{k} \right) \left( b_g + f(k) \mu^2 \right) D_g D_u P_{m \theta}(k).
\end{align}
Our ability to cross-correlate the power spectra of the galaxy velocity field and density field helps to reduce statistical uncertainty in our measurements of the power spectra. This is because any uncertainty in our measurements of the power spectrum of the density field arises due to cosmic variance, and also Poisson noise, because we are essentially randomly sampling galaxies in a field, which introduces a shot noise contribution to the galaxy power spectrum \citep{Peebles1980}. The same can be understood to occur for measurements of the power spectrum of the velocity field, although the velocity power spectrum suffers from additional noise due to the sizeable uncertainty in individual measurements of galaxy peculiar velocities \citep{howlett2017cosmological}. Fortunately, the uncertainties in our measurements of each power spectrum are partially uncorrelated and probe the same underlying fluctuations in the matter field. Hence, cross-correlating the two allows for us to partially cancel the effects of sample variance \citep{Burkey2004,Mcdonald2009}.
\par
\begin{figure}
    \centering
    \includegraphics[scale=0.55]{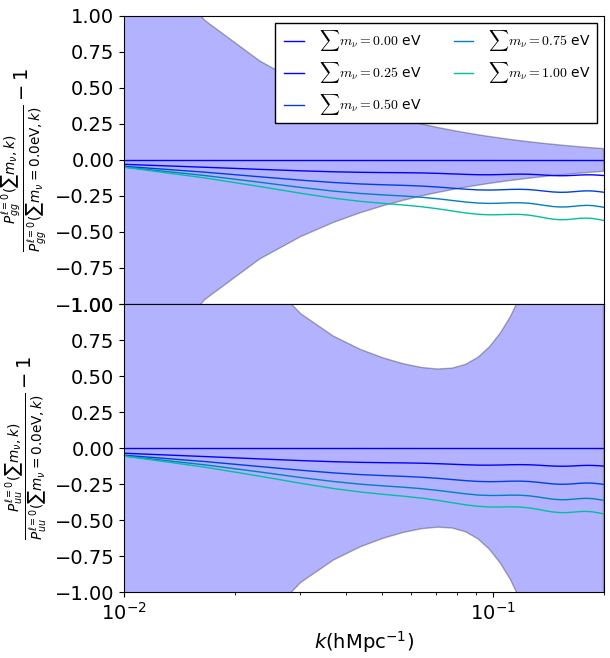}
    \caption{Top panel: ratio of the galaxy-galaxy redshift space power spectrum monopole moment, with varying $\sum m_{\nu}$ to that with $\sum m_{\nu} = 0.0$ eV. Lower panel: the same, except for the velocity-velocity redshift space power spectrum monopole moment. For both cases we have shaded the region of uncertainty for the case $\sum m_{\nu} = 0.0$ eV. The theoretical uncertainty has been calculated for the power spectra moments based on the expected measurements by DESI's Bright Galaxy Survey (BGS) and peculiar velocity survey (see Section~\ref{sec:surveys}), in a redshift bin of width $\Delta z = 0.1$ from $z = 0.05$ to $z = 0.15$, when 30 linear spaced $k$ bins have been measured with $k = 2\times10^{-3}$ to $0.2\,h \mathrm{Mpc}^{-1}$. The error bars have been calculated using the equations in Appendix 2 of \citet{howlett2019redshift}. }
    \label{Figure2_errorbarspowerspectrum}
\end{figure}

 
We can thus expect that measuring both the velocity field and galaxy density field gives more information overall about the underlying matter field and allows for tighter constraints on $\sum m_{\nu}$. Figure~\ref{Figure2_errorbarspowerspectrum} demonstrates that although the typical uncertainties on the velocity power spectrum are much larger than the density power spectrum, there is some accessible information on $\sum m_{\nu}$. While measurements of the power spectra from any individual sample redshift bin may only allow us to constrain $\sum m_{\nu}$ to a few eVs, we expect that a combination of information from multiple power spectra across a full galaxy survey redshift range, or combined with information from measurements of the CMB, can provide more sensitive measurements of $\sum m_{\nu}$. We demonstrate this with Fisher matrix forecasts in Sections~\ref{sec:fisher} and~\ref{sec:results}.

\section{Methods}
\label{sec:fisher}

\subsection{Fisher information}

To answer our questions stated in Section~\ref{intro}, we forecast the ability of future surveys that will sample both galaxy redshifts and PVs to constrain $\sum m_{\nu}$ when they do and do not utilise the PV information. We do this using the Fisher Information method. In this method, the Fisher matrix for a set of free parameters is computed by evaluating the change in some signal with respect to variations in the free parameters given the predicted uncertainties for a given experiment. The inverse of the Fisher information matrix gives the best possible covariance matrix we may obtain for our parameters given the signal. It is important to note that this is based on statistical uncertainty only; the standard Fisher information method does not include sources of systematic error.

\cite{koda2014peculiar} and \cite{howlett2017cosmological} demonstrated that if we treat our observables for a redshift and peculiar velocity survey as the Fourier-space galaxy density and line-of-sight velocity, then the Fisher matrix element for parameters $x_{i}$ and $x_{j}$ is given by
\begin{equation}
    F_{ij} = \frac{\Omega_{\text{sky}}}{4\pi^2} \int_0^{r_{\text{max}}} r^2 dr \int_{k_{\text{min}}}^{k_{\text{max}}} k^2 dk \int_0^1 d\mu \text{ tr}\left( \mathbf{C}^{-1} \frac{d\mathbf{C}}{dx_i} \mathbf{C}^{-1} \frac{d\mathbf{C}}{dx_j} \right).
\end{equation}
Here, $\Omega_{\text{sky}}$ is the sky area of the survey in radians, $\mathbf{C}$ is the covariance matrix for our observables, $\frac{d\mathbf{C}}{dx_i}$ is the derivative of the covariance matrix with respect to the $i^{\mathrm{th}}$ free parameter $x_i$ in our analysis and $r_{\text{max}}$ is the maximum distance that fits inside our redshift bin at $z_{\text{max}}$. $k_{\text{min}}$ is set by $r_{\text{max}}$ and $k_{\text{max}}$ is set by the smallest scale over which we can model the power spectra accurately, usually set to $0.1$ to $0.2\,h \text{Mpc}^{-1}$. The covariance matrix for the galaxy density field and galaxy velocity field is given by \citep{howlett2017cosmological}
\begin{equation}
    \mathbf{C} = \begin{bmatrix}
    P_{gg} + \frac{1}{\bar{n}_g} & P_{ug} \\
    P_{ug} & P_{uu} + \frac{\sigma^2_{\text{total}}}{\bar{n}_u}
    \end{bmatrix},
\end{equation}
with $P_{gg}$, $P_{ug}$, and $P_{uu}$ defined in Section~\ref{sec:PV}. The additional terms on the diagonals represent the shot noise for each power spectrum. $\bar{n}_g$ and $\bar{n}_u$ are the number densities of objects detected as a function of redshift for the galaxy density and velocity surveys respectively, while $\sigma^2_{\text{total}}$ is given by 
\begin{equation}
    \sigma^2_{\text{total}} = \sigma^2_{\text{rand}} + (\alpha H_0 d(z))^2,
\end{equation}
and is the extra contribution to the velocity power spectrum shot noise arising from the typically large uncertainties on individual peculiar velocity measurements. The largest contribution, $\alpha H_0 d(z)$ accounts for total scatter in either the SNe Ia Hubble diagram, the Tully-Fisher relation or the Fundamental Plane relation, where $d(z)$ is the comoving distance to the galaxy and $\alpha$ is the fractional uncertainty in our distance measurement -- typically $\alpha = 0.2$ for the Tully-Fisher relation or Fundamental Plane \citep{Magoulas2012,Howlett2017c,Hong2019}, and $\alpha = 0.05-0.1$ for SNe Ia \citep{rest2014cosmological, fakhouri2015improving}. The second term $\sigma^2_{\text{rand}}=300^{2}\,\text{km}^{2}\text{s}^{-2}$ accounts for non-linear contributions to the peculiar velocities, and ensures that the uncertainty is non-zero even for nearby galaxies.

To incorporate the maximum amount of information that may be obtained by our observables, we also take into account the Alcock-Paczynski (AP) effect \citep{alcock1979evolution}. This is an improvement in the method compared to \cite{koda2014peculiar,howlett2017cosmological,howlett2017measuring}, who neglect this effect given it is relatively small at low redshifts. The AP effect is a geometrical distortion that occurs when one transforms redshift-space coordinates of objects to real coordinates, and assumes a cosmological model that differs from the true cosmology. Essentially, an observer believes they have information on the power spectrum of matter at a certain $k$ and $\mu$ which differs from the real values for $k$ and $\mu$ they are sampling \citep{colas2020efficient}. Deviations of the fiducial cosmological parameters assumed for an analysis on cosmological data, from the `true' cosmology, change the power spectra and the observed $k$ and $\mu$ as a result. Equations relating the fiducial values for $k$ and $\mu$, $k^{\text{Fid}}$ and $\mu^{\text{Fid}}$ to the real $k$ and $\mu$ are given in Appendix \ref{APeq}. Incorporating this into the forecasting analysis allows for all the information that could possibly be obtained from a galaxy survey on $\sum m_{\nu}$ to be included.

\subsection{Free parameters}
\label{sec:freeparameters}

For all forecasts we allow the following cosmological parameters to vary,
\begin{equation}
    \{ \sum m_{\nu}, H_0, \Omega_{\text{b}}h^2, \Omega_{\text{cdm}}h^2, A_s, n_s \}, \label{eq:cosmoparameters} 
\end{equation}
where $H_0=100h\mathrm{km\,s^{-1}\,Mpc^{-1}}$ is the Hubble constant, $\Omega_{\text{b}}h^2$ and $\Omega_{\text{cdm}}h^2$ are the physical densities of cold dark matter and baryons respectively, $A_s$ is the scalar amplitude, which sets the normalization of the linear matter power spectrum, $n_s$ is the spectral index. These parameters are all degenerate with $\sum m_{\nu}$, but constrained to various degrees with galaxy redshift surveys and thus must be allowed to vary for our forecasts. To begin, we fix $N_{\text{eff}}$, the effective number of neutrino species, however we will also present forecasts where $N_{\text{eff}}$ is an additional free parameter in Section~\ref{sec:neff}. We fix $\tau$, the optical depth at reionization, to $\tau = 0.0543$ for our galaxy survey forecasts, which is not constrained by these surveys, but we note that when we combine with data from \textit{Planck} (see Section~\ref{sec:planck}), degeneracies between $\tau$ and other cosmological parameters that are constrained by the CMB are accounted for in the full combined Fisher matrix. 
\par 
All our forecasts use the base TTTEEE+low$\ell$+lensing $\Lambda\text{CDM}$ Planck 2018 results as our fiducial cosmology \citep{aghanim2020planck}, with $N_{\text{eff}} = 3.046$ and $\sum m_{\nu} = 0.058$ eV. We have verified that our forecasts do not change appreciably if we assume a minimal-mass inverted neutrino hierarchy with $\sum m_{\nu} = 0.097$ eV, or if we alter the individual neutrino masses while keeping the sum of masses constant.

\begin{figure*}
    \centering
    \includegraphics[scale=0.51]{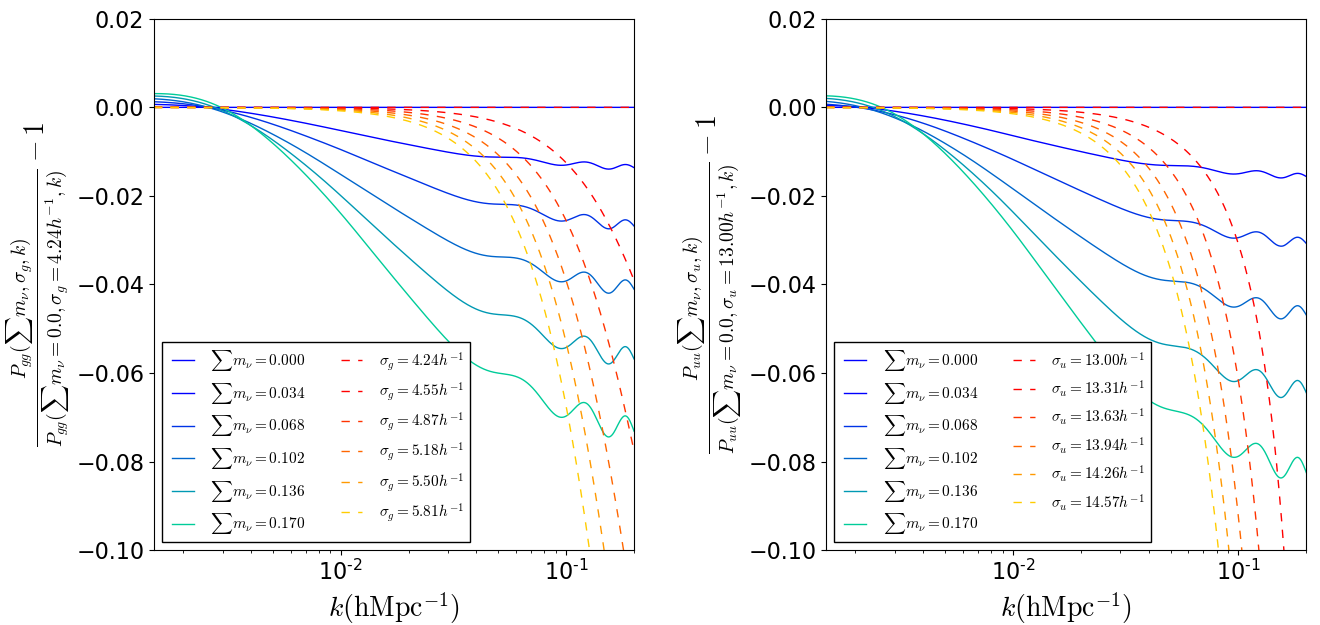}
    \caption{Left: the ratio of the redshift-space galaxy power spectrum with varying $\sum m_{\nu}$ (solid lines) and $\sigma_g$ (dashed lines, $\sigma_g$ is a nuisance parameter associated with damping due to the FoG effect), to the redshift-space galaxy power spectrum with $\sum m_{\nu}$ and $\sigma_g$ fixed. Right: the same, except for the redshift-space velocity power spectrum, and where we vary $\sigma_u$, associated with the damping due to the FoG effect for the velocity power spectrum. $\sigma_g$ and $\sigma_u$ are given in units of Mpc, $\sum m_{\nu}$ in eVs.}
    \label{Figure3_damping_powerspectra}
\end{figure*}

In addition to cosmological parameters, we also include nuisance parameters for each power spectrum forecast. The damping term $D_g$ for the power spectrum of the density field is assumed to take the following functional form,
\begin{equation}
    D_g = \frac{1}{\sqrt{ 1 + \frac{1}{2}(k \mu \sigma_g)^2  }},
\end{equation}
and we allow the unknown constant $\sigma_g$ to vary about a fiducial value of $4.24h^{-1}\mathrm{Mpc}$. This is important as the effect of damping due to the FoG effect is partially degenerate with the effect of damping on the power spectrum due to massive neutrinos; both suppress the galaxy power spectrum on small scales. For each power spectrum the galaxy bias is also treated as a free parameter -- in all cases we parameterize the fiducial galaxy bias following the work of \cite{font2014desi} and the DESI Collaboration Whitepaper \citep{aghamousa2016desi},
\begin{equation}
    b_g(z = 0)D(z = 0) = b_g(z)D(z),
\end{equation}
where $D(z)$ is the linear growth factor at redshift $z$, where $D(z)$ represents the amplitude of a density perturbation. The exact value of $b_{g}(z=0)$ varies for different surveys and our choices for each survey are given in Section~\ref{sec:surveys}. Finally, for the velocity power spectrum of galaxies we introduce one more free parameter. The damping of the velocity power spectrum of galaxies due to non-linear redshift space distortions has been found by \cite{koda2014peculiar} to suit the following functional form,
\begin{equation}
    D_u = \frac{\sin{(k\sigma_u)}}{k\sigma_u}.
\end{equation}
$\sigma_u$ is allowed to vary about $13h^{-1}\mathrm{Mpc}$, which was found by \cite{koda2014peculiar} to reproduce the non-linear clustering of halo velocities in N-Body simulations well. Like $\sigma_{g}$, $\sigma_u$ is also partially degenerate with $\sum m_{\nu}$ and so should be varied. However, in Figure \ref{Figure3_damping_powerspectra}, we can see that although neutrinos and non-linear RSD act on the same parts of the galaxy and velocity power spectra, their functional form is different enough that we expect to still be able to break the degeneracy between the associated parameters. 

The sum of neutrino masses is very small, as is their impact on the matter power spectrum and growth of structure. Hence, in order to obtain meaningful constraints it is necessary to measure the clustering of galaxies over multiple redshifts. For the purposes of our forecasts we split all our surveys into redshift bins of width $\Delta z = 0.1$, allowing the number of galaxies in each bin to vary as would be expected in the real survey. Then, $b_g$, $\sigma_u$ and $\sigma_g$ are treated as \textit{separate} free parameters for each redshift bin we consider. This supposes that we have no knowledge of how the galaxy bias, or the association between galaxies and their cluster/halo properties may evolve with redshift --  a conservative assumption that could potentially be improved upon when measuring cosmological parameters using real galaxy survey data. Overall, the set of nuisance parameters we allow to vary for all of our forecasts includes
\begin{equation}
 \{ b_{g}(z), \sigma_{g}(z), \sigma_{u}(z) \} \label{eq:nuisanceparams}
\end{equation}
where $b_g(z)$ and $\sigma_g(z)$ are included in forecasts that involve redshift data, $\sigma_u(z)$ in forecasts for peculiar velocity data, and all three for forecasts that involve both redshift and peculiar velocity data. Hence, the total free parameters included in our analysis is the set of cosmological parameters in Equation \ref{eq:cosmoparameters} in addition to the relevant nuisance parameters from Equation \ref{eq:nuisanceparams} which depends upon the survey data we perform the analysis on, and which are treated as separate parameters in each redshift bin. This gives the number of total parameters $N$ for a forecast on combined redshift data and peculiar velocity data, when there is $n_{(z,\text{redshift})}$ redshift bins for the redshift data and $n_{(z,\text{velocity})}$ for the peculiar velocity, as
\begin{align}
    N & = N_{\text{cosmo}} + n_{(z,\text{redshift})} N_{(\text{nuisance},\text{redshift})} \nonumber \\ & + n_{(z,\text{velocity})}N_{(\text{nuisance},\text{velocity})}.
\end{align}
When data is combined from separate surveys that involve the same kind of data, the nuisance parameters are still also treated as separate free parameters from each survey dataset we perform the analysis on, because the galaxies can be considered as separate tracers from different surveys which potentially have different underlying values or functional forms for these nuisance parameters. Thus, if we combine data in an analysis from three surveys, where survey 1 contains redshift data, survey 2 contains redshift data and survey 3 contains peculiar velocity data, the above equation would become 
\begin{align}
    N & = N_{\text{cosmo}} + n_{(z,\text{redshift, s1})} N_{(\text{nuisance},\text{redshift, s1})} \nonumber \\
    & + n_{(z,\text{redshift, s2})} N_{(\text{nuisance},\text{redshift, s2})} \nonumber \\ & + n_{(z,\text{velocity, s3})}N_{(\text{nuisance},\text{velocity, s3})}.
\end{align}
In general $N_{(\text{nuisance},\text{redshift})} = 2$ and $N_{(\text{nuisance},\text{velocity})} = 1$ for any single survey, unless there are multiple types of galaxy tracers in the survey data, which is the case for the DESI Baseline survey (see section \ref{sec:desisurvey} for more detail).

\subsection{Numerical Implementation}
\label{sec:numericalimplementation}

For our forecasts we generate the linear matter power spectrum $P_{mm}$, which we assume is equal to $P_{\theta \theta} = P_{m\theta}$, as is true in linear theory, with the \texttt{CLASS} Boltzmann solver \citep{blas2011cosmic}. We compute derivatives $\frac{dP}{dx_i}$ using central finite differences for $x_i$, as $\frac{dP}{dx_i}$ can not be computed analytically for our free cosmological parameters. We tested these derivatives to be robust to different step sizes, and to whether the finite differences were applied to the full redshift space power spectra or to the matter power spectra and growth rate before multiplying by the RSD terms. The step sizes that were found to be robust and were used for each parameter $\Delta x_i$ are given in Appendix~\ref{ndss}. 

For our forecasts, $f(k)$ is computed by redefining the scale-dependent growth factor in terms of the power spectrum itself
\begin{equation}
    f(k) = \frac{d \ln{D(k)}}{d \ln{a}} = \frac{d \ln{P_{mm}^{\frac{1}{2}}}(k)}{d \ln{a}} = \frac{1}{2}\frac{a}{P_{mm}(k)} \frac{dP_{mm}(k)}{da}.
\end{equation}
which makes use of the fact that the linear power spectrum at any reasonably low redshift can be written as a simple rescaling of the redshift zero power spectrum by the ratio of the linear growth factors. The above expression is evaluated using a central finite difference of the matter power spectrum $P_{mm}$ with $\Delta a = 0.0001$, and using a backwards difference about $a = 1$.

The presence of massive neutrinos also introduces a scale dependence to the galaxy bias $b_g$, which is usually assumed to be scale independent on linear scales. However, massive neutrinos introduce a scale dependence to $b_g$ that depends on $\sum m_{\nu}$, even on linear scales \citep{vagnozzi2018bias}, and this makes the bias parameter difficult to model. However we can define the power spectrum, $P_{cb,cb}$, an auto-correlation power spectrum defined with respect to both cold dark matter (c) and baryons (b), in contrast to $P_{mm}$, defined for all matter (cold dark matter, baryons and neutrinos). The bias for this power spectrum, $b_{cb}$, can be treated as a scale independent quantity. Since the effects of redshift space distortions are mainly driven by cold dark matter and baryons, we can just replace $P_{mm} = P_{m\theta} = P_{\theta \theta}$ with $P_{cb,cb}$, in our expression for $f(k)$ and in our expressions for $P_{gg}$, $P_{uu}$ and $P_{ug}$. Likewise we replace $b_g$ with $b_{cb}$ \citep{vagnozzi2018bias}. These substitutions are justified on linear scales. From hereon any references to $b_g$ should be taken to mean the bias defined with respect to baryons and cold dark matter only, and likewise for $P_{cb,cb}$ replacing $P_{mm}$. 

\subsection{Including information from \textit{Planck}}
\label{sec:planck}

Finally, we also add information from the \textit{Planck} 2018 results to our computed Fisher information, in order to obtain forecasts for the ability of surveys with the combination of \textit{Planck} information to constrain $\sum m_{\nu}$. The total information matrix we have will be given by the Fisher information we compute for surveys and an estimate of the Fisher information from \textit{Planck},
\begin{equation}
    \mathbf{F}_{\text{total}} = \mathbf{F}_{\text{surveys}} + \mathbf{F}_{\text{\textit{Planck} 2018}}.
\end{equation}
To estimate the Fisher information from the \textit{Planck} 2018 results, we use the \textit{Planck} 2018 MCMC chains.\footnote{https://wiki.cosmos.esa.int/planck-legacy-archive/index.php/Cosmological Parameters} From these, we construct a covariance matrix for the relevant cosmological parameters, and invert this to obtain $\mathbf{F}_{\text{\textit{Planck} 2018}}$. This process assumes that the likelihood distribution for each parameter is Gaussian, which is clearly not true for $\sum m_{\nu}$ given only upper bounds have been obtained to date; the true contours from \textit{Planck} for $\sum m_{\nu}$ have a hard physical cut off at $\sum m_{\nu} = 0$ eV.

\begin{figure*}
    \centering
    \includegraphics[scale=0.6]{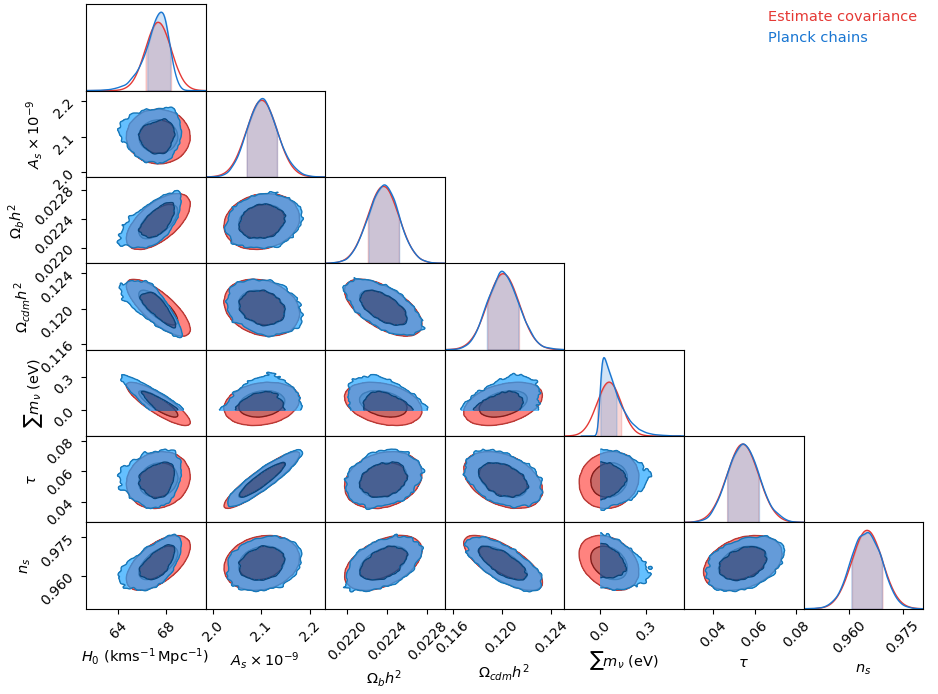}
    \caption{An estimation of the one and two--$\sigma$ contours for cosmological parameters and marginalized
likelihood distributions from the \textit{Planck} 2018 results, based on our estimate of the Fisher
information from \textit{Planck} MCMC chains, using data from the CMB temperature, polarization auto
and cross-correlation power spectra and CMB low$\ell$, lowE and lensing information (in red), vs the actual one and two--$\sigma$ contours for cosmological parameters and marginalized likelihood distributions from the \textit{Planck} 2018 results (blue).}
    \label{Figure4_planckchainscontours}
\end{figure*}

By assuming a Gaussian posterior likelihood distribution for $\sum m_{\nu}$, we are allowing for non-zero probabilities of $\sum m_{\nu}$ < 0. However, we expect that this will make our forecasts conservative when including information from \textit{Planck}, since a hard-cut off at zero would force the likelihood to be greater at values above zero for $\sum m_{\nu}$ and make the error bars we obtain for $\sum m_{\nu}$ smaller. This is because the 68.3\% region of the area under the likelihood distribution curve, about the mean, will have less width. This is demonstrated in Figure~\ref{Figure4_planckchainscontours}. The marginalized likelihood curves for our approximation and the \textit{Planck} chains match very closely for the majority of parameters, but the width of the 1-$\sigma$ region for $H_{0}$ and $\sum m_{\nu}$ is actually slightly larger using the Gaussian approximation compared to the true distribution.

\section{Surveys}
\label{sec:surveys}

\begin{figure}
    \centering
    \includegraphics[scale=0.45]{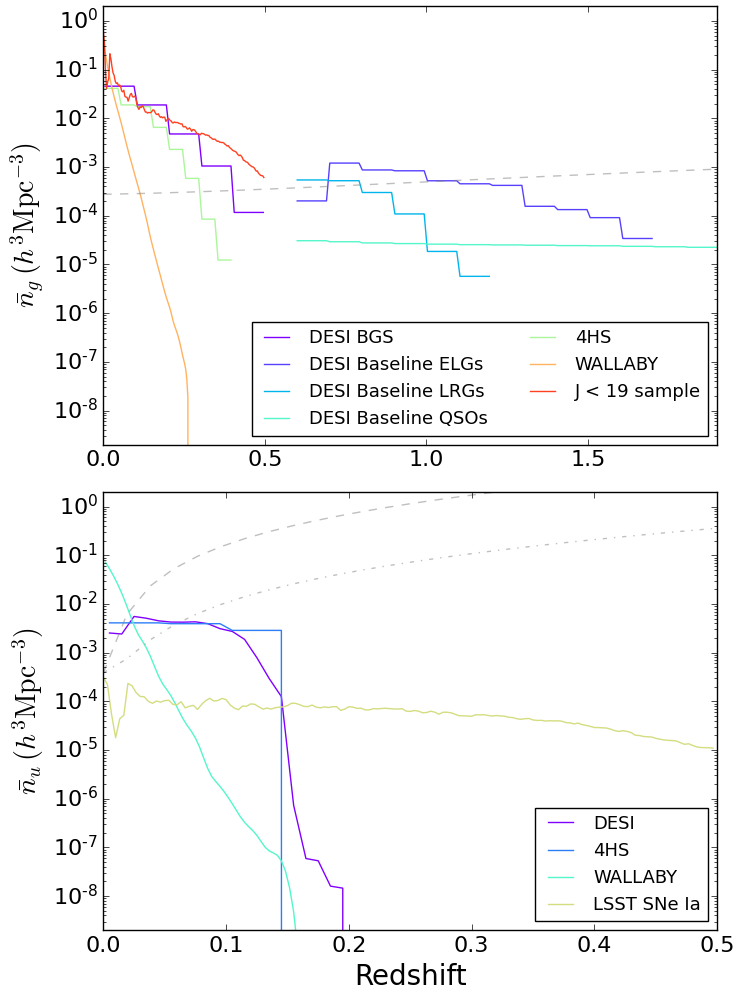}
    \caption{Top panel: the assumed number density of objects for each redshift survey, or separate objects targeted for the DESI Baseline survey. The dashed line shows where the number density of objects is such that the signal to noise ratio $P_{gg}\bar{n}_g$ is unity for a redshift survey. Lower panel: the assumed number densities of objects for each peculiar velocity survey. The dashed line shows where the number density of objects is such that the signal to noise ratio $P_{uu}\frac{\bar{n}_u}{\sigma^2_{\text{total}}}$ is unity for a peculiar velocity survey when the uncertainty on the measured distances to objects to obtain PVs is 20\%, while the dash-dotted line shows the same, but in the case that the uncertainty on the measured distances to objects is 5\%. 
    }
    \label{Figure5_number_density_objects}
\end{figure}

In this section we provide a brief description of each survey we present forecasts for in the results section. Table \ref{table1_surveydetails} summarises the assumed properties for each redshift and peculiar velocity survey, and Figure \ref{Figure5_number_density_objects} shows the number density of objects assumed for each survey. We focus our analysis on WALLABY, 4HS and LSST because these surveys are at low redshift and have already been considered in the literature as demonstrators for the improved constraints on the growth rate of structure offered by combining density and velocity field information. This allows us to now also compare the constraining power of the data from these surveys on $\sum m_{\nu}$ using only redshift data, peculiar velocity data or a combination of both. At high redshift, Both DESI and Euclid are predicted to obtain tight constraints on $\sum m_{\nu}$ in combination with \textit{Planck}. However, in this work we focus only on DESI because DESI will also contain both low redshift data from the Bright Galaxy Survey (BGS) as well as peculiar velocity data from its secondary target program (Saulder et. al., in prep.). Our analysis on the DESI data will allow us to test whether peculiar velocity data is worth pursuing for $\sum m_{\nu}$ constraints, in the scenario that highly constraining data from the DESI survey at high redshift may already already available.

\begin{table}
\caption{Summarised details of survey properties assumed for Fisher forecasts. Peculiar velocities are abbreviated to PVs. }
\renewcommand{\arraystretch}{1.3}
    \centering
    \begin{tabular}{|p{3.5cm}|p{1.0cm}|p{1cm}|p{1.4cm}|} \hline
        \textbf{Survey name} & \textbf{Sky area} ($\text{degrees}^2$) & $\mathbf{b_g(z = 0)}$ (galaxy bias) & \textbf{Distance indicator uncertainty} \\ \hline \hline
        DESI BGS &  14000 & 1.34 & --- \\
        DESI PV survey & 14000 & --- & 20\%  \\
        DESI Baseline survey (ELGs) & 14000 & 0.84 & ---  \\
        DESI Baseline survey (LRGs) & 14000 & 1.7 & ---  \\
        DESI Baseline survey (QSOs) & 14000 & 1.2 & ---  \\
        4HS redshift survey & 17000 & 1.34 & ---   \\
        4HS PV survey & 17000 & --- & 20\%  \\
        WALLABY redshift survey & 20000 & 0.7 & ---    \\
        WALLABY PV survey & 20000 & --- & 20\%   \\
        J $<$ 19 redshift survey & 18000 &  1.34 & ---  \\
        LSST SNe Ia PV survey & 18000 & --- & 5-10\%  \\ \hline
        \label{table1_surveydetails}
    \end{tabular}
\end{table}

\subsection{DESI}
\label{sec:desisurvey}
The DESI (the Dark Energy Spectroscopic Instrument) is a multi-fiber optical spectrograph that is mounted on  the  Kitt  Peak  National Observatory Mayall 4m telescope that began a 5-year redshift survey of galaxies and quasars in  2021.\footnote{https://www.desi.lbl.gov/} DESI is expected to obtain redshifts for 34 million galaxies. The DESI collaboration Whitepaper outlines that DESI's Baseline survey will target emission line galaxies (ELGs), luminous red galaxies (LRGs) and quasi-stellar objects (QSOs), also referred to here as quasars \citep{aghamousa2016desi}. When the moonlight is bright and does not allow for faint targets to be detected, DESI will conduct a magnitude limited bright galaxy survey (BGS) that will detect low redshift galaxies \citep{aghamousa2016desi}. 
\par 
The number density of objects that we assume will be detected during the DESI redshift surveys is taken from the numbers predicted by the DESI Collaboration whitepaper \citep{aghamousa2016desi}. We have used a cosmological model consistent with our forecasts to convert the number of objects per redshift bin $\frac{dN}{dz}$ given in the paper to the number of objects per $h^{-3}\text{Mpc}^{3}$. The DESI Whitepaper states their redshift surveys will cover $14,000 \,\mathrm{deg}^{2}$ of the sky, which we also use for our forecasts. For BGS, we assume $D(z=0)b_g(z = 0) = 1.34$, as in the whitepaper. For the Baseline survey, the whitepaper outlines that different galaxy biases are used for different tracers (ELGs, QSOs, LRGs) and the forecasts they provide for different tracers are cross-correlated. We assume the same biases as outlined in the whitepaper for each tracer and similarly cross-correlate the information for different tracers.
\par 
We also provide forecasts for a DESI peculiar velocity survey, which has recently begun at the time of writing using spare fibres on the DESI instrument (Saulder et. al., in preparation). For this survey, we assume data is obtained by measuring distances to galaxies detected in the BGS redshift sample using the Fundamental plane relation for early-type galaxies (although the survey will also contain a large number of Tully-Fisher based distances). Measurements of peculiar velocities using these tracers have an approximate uncertainty of 20\% on the distances to galaxies \citep{Magoulas2012}, which we assume for the forecasting analysis. The number densities of objects expected to be detected are estimated using photometric cuts applied to the DESI Legacy Imaging Survey data \citep{Dey2019}, and assuming that $\sim$ 50\% of the targets are unsuitable for use in the Fundamental Plane once spectra have been taken (e.g., due to the presence of H$\alpha$ in the spectrum). The total number of predicted galaxies in this sample is $\sim220,000$; a substantial increase over the largest currently available sample (The SDSS Peculiar Velocity catalogue containing $\sim34,000$ galaxies; Howlett et. al., in preparation) due to it going a full magnitude deeper, over twice the sky area, and to slightly higher redshift. We assume the density field and velocity field from the DESI BGS and peculiar velocity survey cover the same area of sky and so are fully overlapping. Overall, we provide forecasts for the BGS redshift sample alone, the Baseline redshift sample alone, the DESI peculiar velocity sample alone, and combinations of these surveys together.

\subsubsection{Forecasting for the DESI Baseline survey}

Due to the presence of three overlapping tracers of the density field in the DESI Baseline survey, we require a slight modification to the method described in Section~\ref{sec:fisher}. Our free parameters include a bias parameter and a damping parameter $\sigma_g$ for each separate tracer in each redshift bin. Our covariance matrix for the three separate density fields, $\delta_{\text{ELG}}$, $\delta_{\text{LRG}}$ and $\delta_{{QSO}}$, becomes

\begin{equation}
    \mathbf{C} = \left[ \begin{matrix}
    P_{{\text{ELG}}} + \frac{1}{\bar{n}_{\text{ELG}}} & P_{{\text{ELG-LRG}}} & P_{{\text{ELG-QSO}}} \\
    P_{{\text{ELG-LRG}}} & P_{{\text{LRG}}}  + \frac{1}{\bar{n}_{\text{LRG}}} & P_{{\text{LRG-QSO}}} \\
    P_{{\text{ELG-QSO}}} & P_{{\text{LRG-QSO}}} & P_{{\text{QSO}}}  + \frac{1}{\bar{n}_{\text{QSO}}} \\
    \end{matrix} \right].
\end{equation}

It can be understood that the auto-power spectrum of the tracer $x$, $P_{xx}$, is given by
\begin{equation}
    P_{xx} = \left( b_{x} + f(k) \mu^2\right)^2 D^2_{g_x} P_{mm},
\end{equation}
and the cross power spectrum of the tracers $x$ and $y$ is given by
\begin{equation}
    P_{xy} = \left( b_{x} + f(k) \mu^2\right)\left( b_{y} + f(k) \mu^2\right) D_{g_x} D_{g_y} P_{mm}.
\end{equation}
$\bar{n}_{\text{x}}$ is the number density of objects for tracer $x$, $b_x$ is the linear bias relative to the underlying matter distribution associated with the objects for tracer $x$ and $D_{g_x}$ is the damping term due to the FoG effect associated with the tracer $x$. Although in practice we set the fiducial value of $\sigma_g$ to be the same for each tracer, writing the spectra in this way allows us to ensure that conceptually we treat them as independent free parameters.

\subsection{WALLABY}

The Widefield ASKAP L-band Legacy All-sky Blind surveY (WALLABY) is currently running at the CSIRO ASKAP (Australian Square Kilometre Array Pathfinder) radio telescope in Western Australia, at the Murchison Radio-astronomy observatory.\footnote{https://wallaby-survey.org/overview/} This telescope will detect and map the 21-cm line of neutral hydrogen to map the positions of galaxies and the large scale structure (LSS) of the Universe. WALLABY will cover approximately three-quarters of the sky over two years to a redshift of $z \approx 0.26$ \citep{koribalski2020wallaby}. WALLABY is expected to detect approximately half a million galaxies \citep{koribalski2020wallaby}. \par 
The number densities used for forecasts for this survey are computed by \citet{howlett2017cosmological}, which are based on the specifications of the survey and simulations by \citep{duffy2012predictions}. 
With this selection function, $\sim324,000$ galaxies will be detected by WALLABY and peculiar velocities may be obtained for $\sim17,000$ of these objects, using the Tully-Fisher relation (and thus a 20\% uncertainty for distance indicator measurements is assumed \citealt{Hong2019}). We adopt the same galaxy bias of $b_g = 0.7$ and sky area of $20,000 \,\mathrm{deg}^{2}$ used by these authors.

\subsection{4HS}

The 4MOST (4-metre multi-object spectrograph telescope\footnote{https://www.eso.org/public/teles-instr/paranal-observatory/surveytelescopes/vista/4most/}) hemispheric survey (4HS; PIs Cluver and Taylor) is a proposed low-redshift future survey on the 4MOST telescope that will map millions galaxies in the southern hemisphere up to redshift of $z \approx 0.15$ \citep{taylor2020taipan} and produce both redshift and peculiar velocity measurements.

The number densities we use for forecasts are taken from the proposal, and are based on the expected distribution of galaxies from the Dark SAGE semi-analytic galaxy formation model by \citep{stevens2017dark}. The 4HS redshift sample will be limited to observing galaxies with $J < 18$ and also $J - K < 0.45$, resulting in a target density of $\sim325$ galaxies per deg$^{2}$. We also assume that 4HS will obtain peculiar velocities using the Fundamental plane relation for 450,000 galaxies (about $8\%$ of the redshift sample), with a 20\% uncertainty for the measurements of distances to objects with this relation. This is larger than the predicted number of the DESI PV survey as the sample has been designed with PV studies in mind, rather than using spare fibres, and uses a longer exposure time which returns a higher rate of successful velocity dispersion measurements. We assume a sky area of $17,000 \,\mathrm{deg}^{2}$ and adopt a galaxy bias of $b_g = 1.34$. We choose this since it is expected that the 4HS redshift sample will be functionally similar to the DESI BGS, discussed previously, and the $J < 19$ sample discussed in the next section. 

\subsection{LSST and \textit{J} < 19 sample}
The Vera C. Rubin Observatory is currently under construction in Chile, and will soon begin the Legacy Survey of Space and Time (LSST).\footnote{https://www.lsst.org/about} It is expected to be operational and running by 2023 \citep{ivezic2019lsst}. The `Wide, Fast and Deep' aspect of this survey will run for approximately 10 years and detect up to 3-4 million SNe Ia, some of which may be used to determine galaxy peculiar velocities. We present forecasts here with an assumed uncertainty of 10\% on distances to galaxies from SNe Ia, as well as optimistic forecasts for when the distance measurements to SNe Ia host galaxies have as little as 5\% uncertainty. \par 
We use the predicted number density of peculiar velocities measured from LSST SNe Ia calculated by \citet{howlett2017measuring}. These predictions are based on the number of SNe Ia that we could expect to be detected by LSST and that will also have also host galaxy redshifts, where the availability of host redshifts for SNe Ia is determined based on a hypothetical $J < 19$ redshift sample in the LSST footprint. The number of host redshifts in this hypothetical sample is predicted using a Semi-Analytic galaxy simulation. Although hypothetical, Section 3.2.2 of \cite{howlett2017measuring} demonstrates that the expected target density ($\sim900\,\mathrm{deg^{-2}}$) would be comparable to DESI BGS \citep{Ruiz-Macias2020}. It is approximately one magnitude deeper than the above proposed 4MOST Hemisphere Survey, but could be considered as either a rough combination of this and other confirmed 4MOST Consortium surveys \citep{deJong2019,Driver2019,Finoguenov2019,Merloni2019,Richard2019}, or a possible extension to these towards the end of the LSST timeline. Overall, the authors predict that approximately $160,000$ SNe Ia could be detected by LSST that will have host redshifts from the overlapping redshift sample.
\par 
We provide forecasts for the constraining power of the $J < 19$ sample alone, the LSST peculiar velocity sample alone, and the combination of the information from these samples, when the total sky area is $18,000 \,\mathrm{deg}^{2}$ and for when $b_g(z = 0) = 1.34$, as in \citet{howlett2017measuring}. The assumed sky area is based on the LSST observing strategy \citep{marshall2017science} and the assumed galaxy bias for the $J < 19$ sample is based on the bias for DESI BGS \citep{aghamousa2016desi}, which will detect a similar number density of objects in the same redshift range.

\section{Results}
\label{sec:results}

\subsection{Results with \texorpdfstring{$N_{\text{eff}}$}{t} fixed}

In this section we present our forecasts for the constraints on $\sum m_{\nu}$ when $N_{\text{eff}}$ is kept fixed. We begin with a comparison of individual surveys as described in Section~\ref{sec:surveys}, before considering combinations that cover a wider redshift range.

\subsubsection{Individual surveys}

\begin{table}
    \caption{Forecasts for surveys where an estimate of the Fisher information from the \textit{Planck} 2018 results is included, for $k_{\text{max}} = 0.1h (0.2h) \mathrm{Mpc}^{-1}$. $\alpha$ gives the assumed fractional uncertainty on SNe Ia for these forecasts (we have provided forecasts for two assumed values of $\alpha$ for LSST SNe Ia) - the assumed fractional uncertainty on distance indicators for PV survey forecasts, other than LSST, are as specified in Table \ref{table1_surveydetails}. The second column in this table shows the predicted uncertainty on $\sum m_{\nu}$, $\sigma_{ \sum m_{\nu} }$. The third column shows the relative improvement when adding information from the PV survey (abbreviated red+PV) compared to the corresponding redshift-only survey (red). 
    }
    \centering
    \begin{tabular}{p{3.9cm}|c|c}
    \hline
    \textbf{Surveys} (+Planck 2018) & $\sigma_{\sum m_{\nu}}$ (eV) & 
    $1 - \frac{\sigma_{\text{red+PV}}}{ \sigma_{\text{red}}} \left( \% \right)$ \\ \hline \hline
    DESI BGS & 0.071 (0.060) & -- \\ 
    DESI PVs & 0.077 (0.077) & -- \\ 
    DESI (BGS + PVs) & 0.070 (0.059) & 1.4 (1.7) \\ 
    WALLABY redshifts &  0.077 (0.074) & -- \\ 
    WALLABY PVs &  0.077 (0.077) & -- \\ 
    WALLABY (redshifts and PVs) &  0.069 (0.067) & 10.4 (9.5) \\
    4HS redshifts &  0.074 (0.064) & -- \\ 
    4HS PVs & 0.077 (0.077) & -- \\ 
    4HS (redshifts and PVs) & 0.072 (0.063) & 2.7 (1.5) \\ 
    $J < 19$  & 0.068 (0.056) & -- \\
    LSST PVs ($\alpha = 0.05$) & 0.077 (0.077) & -- \\ 
    LSST PVs ($\alpha = 0.1$) & 0.077 (0.077) & -- \\ 
    $J < 19$ + LSST PVs ($\alpha = 0.05$) &  0.068 (0.055) & 0 (1.8)\\ 
    $J < 19$ + LSST PVs ($\alpha = 0.1$) & 0.068 (0.056) & 0 (0) \\
    DESI Baseline &  0.050 (0.035) & -- \\
    \hline 
\end{tabular}
\label{results_table_1_neff_fixed}
\end{table}

Our forecasts for individual surveys are shown in Table~\ref{results_table_1_neff_fixed} and are able to obtain improvements between 0\% -- 10\% for $k_{\text{max}} = 0.1h \mathrm{Mpc}^{-1}$ for the surveys we consider. The fact that the improvement is not always zero implies that PV surveys do contain some information ob $\sum m_{\nu}$.  In the case $k_{\text{max}} = 0.2h \mathrm{Mpc}^{-1}$, the percent improvement is similar.

For $k_{\text{max}} = 0.2h \mathrm{Mpc}^{-1}$, the improvement is slightly less when adding the PV data to the 4HS and WALLABY surveys, compared to the $k_{\text{max}} = 0.1h \mathrm{Mpc}^{-1}$ case. For DESI BGS, and the $J < 19$ survey where $\alpha = 0.05$, the improvement is actually greater when $k_{\text{max}} = 0.2h \mathrm{Mpc}^{-1}$, although it is zero for the $J < 19$ sample when $\alpha = 0.1$. A weaker improvement on constraints when $k_{\text{max}} = 0.2h \mathrm{Mpc}^{-1}$ in the case of WALLABY or 4HS PVs can be explained because the inclusion of more non-linear scales in the measurement of the density field means the redshift surveys probe more of the signal due to the impact of neutrino free-streaming, which becomes larger on smaller scales, and because there are also more unique modes for us to measure, which reduces the uncertainty. Both of these factors allow for tighter constraints from redshifts alone, but less so when adding data from the velocity field. This is because the presence of uncertainties on the measurements of the velocity field increases the shot-noise at these wavelengths. Ultimately, the signal-to-noise ratio at more non-linear wavelengths will dictate how much more improvement is possible when including PV measurements.

\begin{figure*}
    \centering
    \includegraphics[scale=0.6]{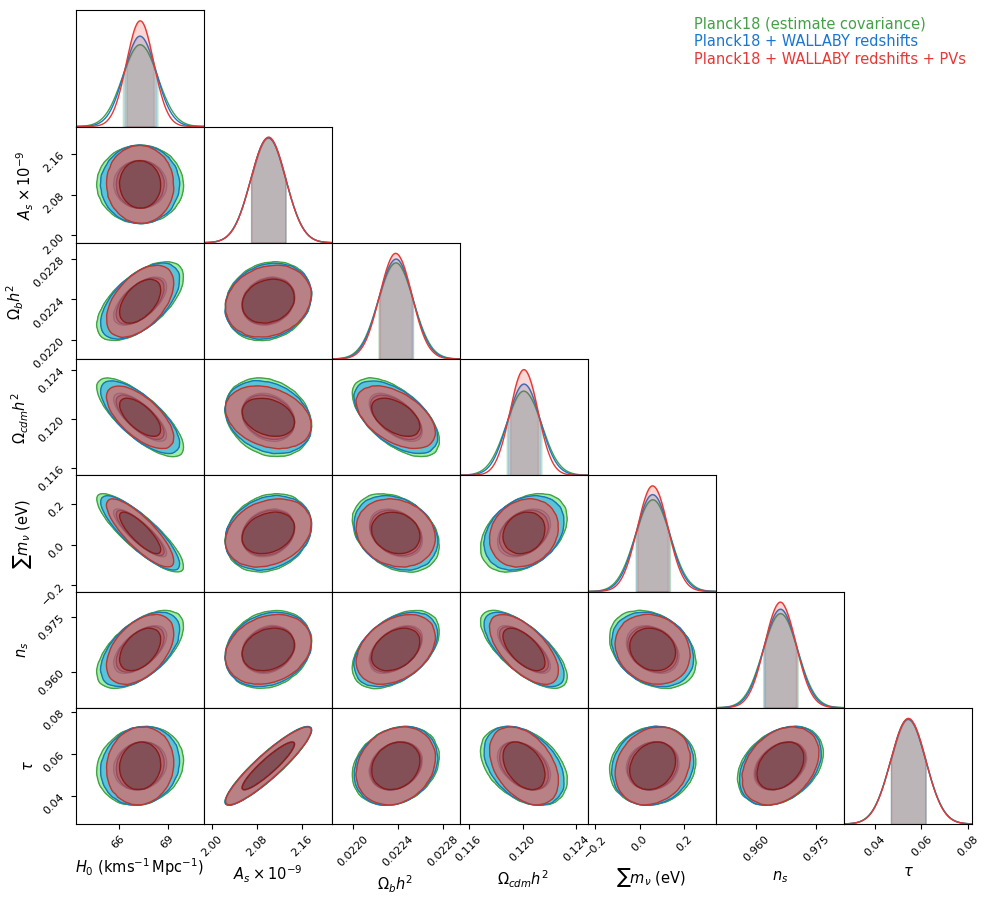}
    \caption{1 and 2-$\sigma$ contours and marginalised likelihoods for cosmological parameters, for our estimate of the Fisher information from the \textit{Planck} 2018 data in green, then the contours when we add the Fisher information from the WALLABY redshift survey for $k_{\text{max}} = 0.2 h \mathrm{Mpc}^{-1}$ (blue contours), and then with the further addition of the WALLABY PV survey information, also for $k_{\text{max}} = 0.2 h \mathrm{Mpc}^{-1}$ (red contours). We also show contours for $\tau$ although $\tau$ is not included as a free parameter in our forecasting analysis, because there is information on $\tau$ from the \textit{Planck} data, which is degenerate with varying $\sum m_{\nu}$ for CMB data. The darker red contours represent 1-sigma constraints for all of the combined datasets and pink represent the 2-sigma constraints. 
    }
    \label{Figure6_contours_neff_fixed}
\end{figure*}

Figure~\ref{Figure6_contours_neff_fixed} shows the 1 and 2--$\sigma$ contours for the free cosmological parameters in our forecasts for the WALLABY redshift survey combined with the WALLABY PV survey information and with our estimate of the \textit{Planck} information. Figure~\ref{Figure6_contours_neff_fixed} shows that the addition of the WALLABY peculiar velocity information to the WALLABY redshifts and \textit{Planck} is able to reduce the degeneracy between $\sum m_{\nu}$ and $H_0$, $\Omega_{\text{cdm}}h^2$, $\Omega_{\text{b}}h^2$, $A_s$ and $n_s$. Any improvement in constraints when adding PV survey information can be attributed to the fact that the redshift-space velocity power spectrum contains the same matter power spectrum and growth rate of structure (and impact on these due to massive neutrinos) that the redshift-space galaxy power spectrum contains, while not suffering from galaxy bias. Furthermore, both redshift and peculiar velocity surveys add to the \textit{Planck} information on cosmological parameters which are degenerate with $\sum m_{\nu}$, particularly those that affect the late time expansion/evolution of the Universe such as the Hubble constant and $\Omega_{\text{cdm}}h^{2}$.

The results tell us that the velocity field \textit{does} contain some information on the sum of neutrino masses, that can be accessed with current/next-generation surveys and can improve constraints. Overall, the improvement is most substantial when the relative shot noise of the peculiar velocity field is low (which can be most easily achieved by increasing the number density at low redshifts) and in some cases when the analysis is limited to quasi-linear scales.

Notably the best forecast in Table \ref{results_table_1_neff_fixed} is given by the DESI Baseline survey for $k_{\text{max}} = 0.2h \mathrm{Mpc}^{-1}$ which only includes redshifts. This then leads into our second question -- even though the velocity field contains useful information at low redshift, is this worth the effort when compared to the full range of information that a typical redshift survey provides? We expect that data from the DESI Baseline survey, which will detect galaxies at high redshift, will be able to obtain a tight constraint on $\sum m_{\nu}$. Thus in the next section we focus our analysis on adding information from peculiar velocities and other low redshift galaxy surveys to the DESI Baseline survey, to test the worthiness of pursuing peculiar velocities in the context of obtaining $\sum m_{\nu}$ constraints.

\subsubsection{Combined surveys}

\begin{table*}
    \caption{Forecasts for constraints on $\sum m_{\nu}$ for different combinations of datasets, for $k_{\text{max}} = 0.1h (0.2h) \mathrm{Mpc}^{-1}$. Pairs of columns show results with and without the inclusion of \textit{Planck}. $\alpha$ gives the assumed fractional uncertainty on the distances to SNe Ia used for LSST PV forecasts. For other PV surveys, Table \ref{table1_surveydetails} specifies the assumed uncertainty on the distance indicator for PVs. The second and fourth columns show the predicted uncertainty on the sum of neutrino masses, while the third and fifth columns show the relative improvement on the uncertainty for each combination compared to the DESI Baseline survey (DESI BL) alone. 
    }
    \label{corrected_version_table3}
    \centering
    \renewcommand{\arraystretch}{1.3}
    \begin{tabular}{p{8.cm}|c|c|c|c}
    \hline
    \multirow{2}{*}{\textbf{Surveys}} & \multicolumn{2}{c}{\textbf{With \textit{Planck}}} & \multicolumn{2}{c}{\textbf{Without \textit{Planck}}} \\
    & $\sigma_{\sum m_{\nu}}$ (eV) & $1 - \frac{\sigma}{ \sigma_{\text{DESI BL}}} \left(\%\right)$ & $\sigma_{\sum m_{\nu}}$ (eV) & $1 - \frac{\sigma}{ \sigma_{\text{DESI BL}} } \left(\%\right)$ \\ \hline \hline
    DESI BL & 0.050 (0.035) & -- & 0.459 (0.216) & -- \\
    DESI BL + BGS & 0.048 (0.034) & 4.0 (2.9) & 0.407 (0.188) & 11.3 (13.0) \\
    DESI BL + BGS + DESI PVs & 0.048 (0.034) & 4.0 (2.9) & 0.34 (0.156) & 26.0 (27.8) \\
    DESI BL + BGS + $J < 19$ & 0.047 (0.033) & 6.0 (5.7) & 0.362 (0.161) & 21.1 (25.5) \\
    DESI BL + BGS + DESI PVs + $J < 19$ & 0.046 (0.033) & 8.0 (5.7) & 0.314 (0.139) & 31.6 (35.6) \\
    DESI BL + BGS + DESI PVs + LSST PVs ($\alpha = 0.05$) & 0.048 (0.034) & 4.0 (2.9) & 0.331 (0.150) & 27.9 (30.6) \\
    \hline 
\end{tabular}
\end{table*}

Table~\ref{corrected_version_table3} shows the constraints for combinations of surveys with the DESI Baseline survey, with and without the inclusion of \textit{Planck} information. For the results that include our estimate of \textit{Planck} information, when $k_{\text{max}} = 0.2 h \mathrm{Mpc}^{-1}$, the addition of either DESI peculiar velocities or BGS data has \textit{negligible} impact on the constraint for $\sum m_{\nu}$ compared to the DESI Baseline redshifts. When information on smaller scales is limited ($k_{\text{max}} = 0.1 h \mathrm{Mpc}^{-1}$), we see a small improvement of $\sim 4-8\%$ by adding information from some surveys. The addition of DESI BGS does improve the constraint very slightly, but the addition of DESI peculiar velocities provides no further improvement. Adding more peculiar velocities from LSST appears to makes little further improvement for either choice of $k_{\text{max}}$, compared to the improvement of including information from the $J < 19$ redshift survey.

All of this is to say that, in general, when considering the combination of \textit{Planck} and DESI Baseline, the addition of low redshift data (density or velocity field) is not particularly helpful; the additional cosmological volume is small compared to the already enormous portion of the Universe covered by simply going to higher redshifts. The PV information provides no improvement to our neutrino mass constraints from DESI Baseline + BGS + \textit{Planck} in any case we consider.

However, the story is quite different when we consider the results \textit{without} \textit{Planck} information. The improvement with the addition of DESI PV information to the two DESI redshift surveys, BGS and the Baseline survey, is notable. The uncertainty on $\sum m_{\nu}$ is reduced by $\sim 15\%$ for $k_{\text{max}} = 0.1 h \mathrm{Mpc}^{-1}$, and $\sim 17\%$ when $k_{\text{max}} = 0.2 h \mathrm{Mpc}^{-1}$. However, if we compare the results for this scenario, to the result for the improvement to DESI Baseline+BGS by adding more information from the $J < 19$ redshift sample, we gain only an improvement of $\sim 11-14\%$ for either $k_{\text{max}}$ choice. This shows that there are some scenarios where a high density peculiar velocity sample is more effective at improving the constraint on $\sum m_{\nu}$ than adding another redshift survey. Essentially, adding a high density peculiar velocity sample is more successful at overcoming the sample variance limit than covering a larger sky area with only redshifts. 

It is interesting to consider whether combinations of these redshift and PV surveys alone are enough to obtain strong constraints that are independent of early-Universe probes in order to compare late and early-Universe constraints. The best constraint we obtain with our results in Table \ref{corrected_version_table3}, from combinations of PV surveys with redshift surveys is $\sum m_{\nu} = 0.139$ eV, and for the scenario we only include DESI surveys, DESI Baseline+BGS+DESI PVs, we may obtain $\sum m_{\nu} = 0.156$ eV. Overall, for the surveys we consider, our forecasts do not suggest it is possible to use upcoming galaxy surveys to obtain a constraint on $\sum m_{\nu}$ that is independent of CMB information by using the peculiar velocity information and also comparable to \textit{Planck} constraints, for the surveys we consider. To obtain a constraint for $\sum m_{\nu}$ that is independent of \textit{Planck} or early-Universe probes will require including additional information from other late-Universe probes. However, overall PV surveys are clearly still able to aid in improving $\sum m_{\nu}$ constraints that are independent of early-Universe probes. 

\begin{figure*}
    \centering
    \includegraphics[scale=0.43]{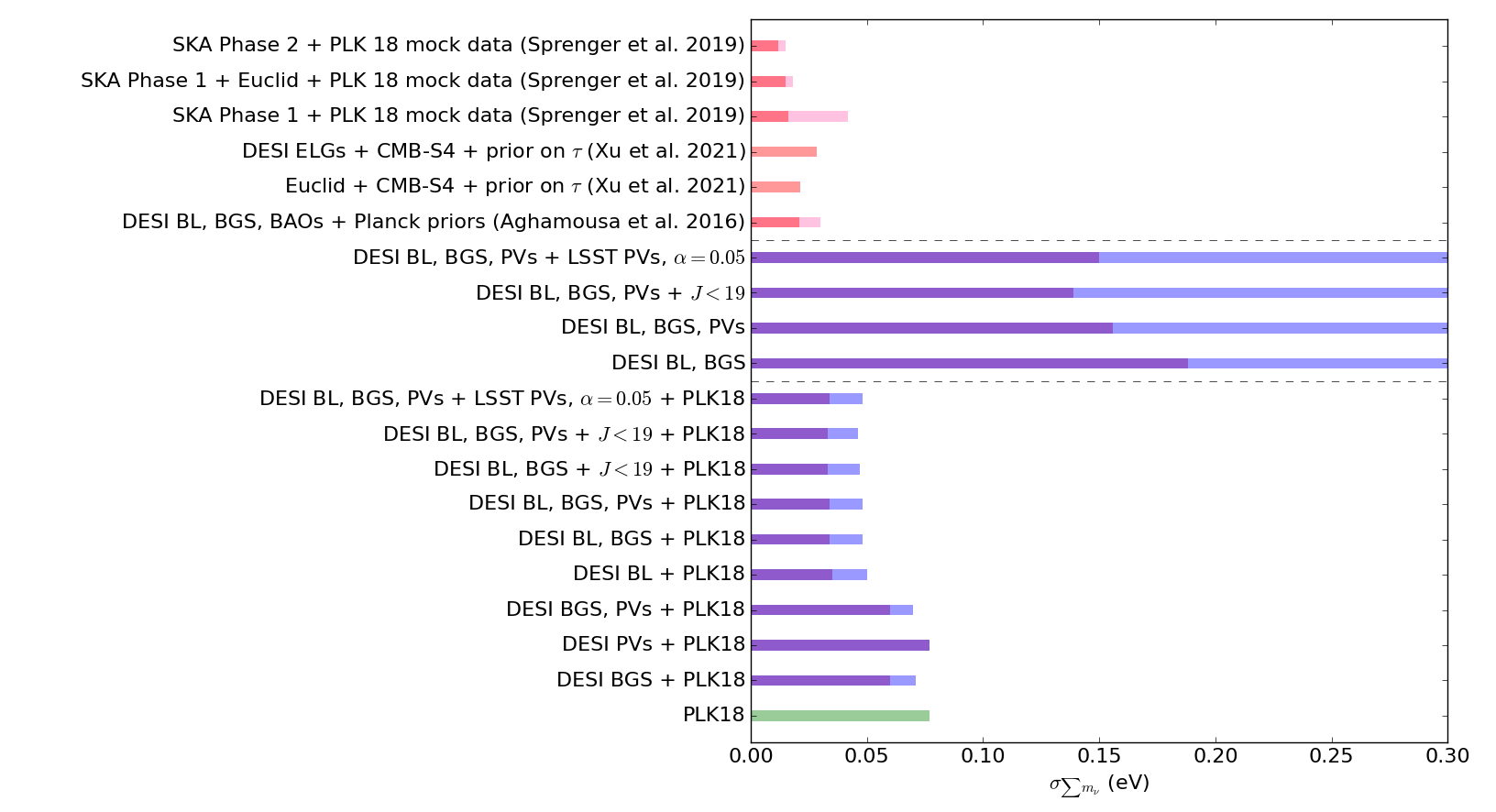}
    \caption{This plot compares forecasts from the corrected results of this work to our estimate of the Fisher information in the \textit{Planck} 2018 results (green), combinations of our forecasts with or without our estimate of the \textit{Planck} information (blue and purple, for $k_{\text{max}} = 0.1h \mathrm{Mpc}^{-1}$ and $k_{\text{max}} = 0.2h \mathrm{Mpc}^{-1}$ respectively), where forecasts that do not include the fisher information from \textit{Planck} are between the two dashed lines, and finally forecasts from other works (red bar or for two forecasts, a red and pink bar), shown above the upper dashed line. Information from the DESI Baseline survey is abbreviated to DESI BL, peculiar velocities to PV, our estimate of the \textit{Planck} 2018 TTTEEE + low$\ell$ + lowE + lensing information to PLK18. For the  \protect\cite{aghamousa2016desi} forecast, the pink (red) bar corresponds to $k_{\text{max}} = 0.1h \mathrm{Mpc}^{-1}$ ($k_{\text{max}} = 0.2h \mathrm{Mpc}^{-1}$). For forecasts by \protect\cite{sprenger2019cosmology}, the pink (red) bars correspond to `conservative' ('realistic') forecasts. Only one forecast (red) is shown for different data combinations by \protect\cite{xu2021accurately}. Where the bars go off the plot we have not shown the full range, because it takes away from the small differences that can be seen in the other results, but the reader may assume the values of the uncertainties shown are significantly larger than $0.3$ eV (refer to Table \ref{corrected_version_table3} for numerical values). 
    }
    \label{results_barchart_nefffixed}
\end{figure*}

\subsubsection{Comparison to other works \label{sec::otherworks}}
 
Figure~\ref{results_barchart_nefffixed} summarises the most interesting results discussed in this section and shown in Tables \ref{results_table_1_neff_fixed} and \ref{corrected_version_table3}. It also shows a comparison of these results to forecasts by three other works; \citet{aghamousa2016desi, sprenger2019cosmology, xu2021accurately}. 
\par 
Comparing our results to the DESI Collaboration's Whitepaper $\sum m_{\nu}$ forecasts \citep{aghamousa2016desi} for the Baseline survey with BGS and BAO data in Figure \ref{results_barchart_nefffixed}, the results are in reasonably good agreement given the differences in modelling -- our constraints for DESI Baseline + BGS + \textit{Planck} are weaker for both choices of $k_{\text{max}}$ by roughly a factor of $\sim 1.7$ than those of \cite{aghamousa2016desi}. However, we do not include BAO information beyond $k_{\text{max}}$ and we also marginalise over nuisance parameters in our modelling of non-linear RSD, which are partially degenerate with the sum of neutrino masses. Furthermore, we include the information from \textit{Planck} 2018 data rather than \textit{Planck} 2013 forecasts, which are included in the form of priors in the work of \cite{aghamousa2016desi}, so these differences seem reasonable.
\par 
We also compare the forecasts here to some of those by \citet{sprenger2019cosmology}, for the Square kilometre array (SKA), and Euclid, with mock \textit{Planck} 2018 data. Euclid is expected to be launched in 2022, and will measure billions of galaxy redshifts out to $z \sim 2$, and will measure weak gravitational lensing.\footnote{https://sci.esa.int/web/euclid/-/summary} The SKA\footnote{https://www.skatelescope.org/} is an array of telescopes that will have a total combined collecting area of approximately 1 km$^2$, with arrays of telescopes located in Australia and South Africa. SKA is expected to begin observing in the late 2020s, and will map the distribution of hydrogen in the Universe and conduct weak lensing galaxy surveys. The red bars and pink bars in Fig.~\ref{results_barchart_nefffixed} correspond to `realistic' forecasts and `conservative' forecasts for these datasets, respectively (a greater sky volume is probed in the `realistic' forecasts, information is obtained on more non-linear scales for intensity mapping, cosmic shear and galaxy clustering).
\par 
Our results for the DESI surveys with \textit{Planck} have overall weaker constraining power compared to forecasts for these data combinations by \cite{sprenger2019cosmology} for any of the constraints we compare, by a factor of $\sim 2$. Both the SKA and Euclid will cover a large cosmological volume like the DESI surveys. SKA Phase 1, covers a smaller sky area ($\sim5000 \text{deg}^2$), but this is compensated for by the additional information provided by intensity mapping at high redshift. SKA Phase 2 is expected to cover 30,000 $\text{deg}^2$ and Euclid about 15,000 $\text{deg}^2$, so equal to or greater than DESI, but again will benefit from additional constraints from multiple cosmological probes such as intensity mapping and weak lensing. These factors explain the better constraints on $\sum m_{\nu}$ predicted for these surveys. We expect the SKA and Euclid are unlikely to be complimented by peculiar velocity measurements in the context of obtaining better constraints for $\sum m_{\nu}$, when \textit{Planck} information is included. It could potentially be interesting to consider the usefulness of PVs when information from \textit{Planck} is \textit{not} included in addition to the SKA and Euclid data, as we found that peculiar velocities improved our forecasts strongly for the DESI redshift surveys alone -- particularly because the SKA could produce a large number of Tully-Fisher peculiar velocity measurements itself and build substantially on the work of WALLABY, one of its precursor surveys.
\par 
We finally compare our results to forecasts for Euclid and DESI ELGs (the same number density of objects is assumed as for our own forecasts for DESI ELGs) with data from CMB stage 4 measurements (CMB-S4), produced by \citet{xu2021accurately}. CMB-S4 refers to a collection of several ground-based telescopes that will measure the CMB over a period of seven years at two locations; the South Pole and the Atacama Plateau in Chile; these telescopes will be equipped with sensitive superconducting cameras which will allow for more precise measurements of the CMB than past experiments.\footnote{https://cmb-s4.org/overview.php} \cite{xu2021accurately} assume the same survey properties for the DESI Baseline survey as for our own forecasts and the DESI Collaboration Whitepaper \citep{aghamousa2016desi}, and the same properties as \cite{sprenger2019cosmology} for Euclid, although the authors do not specify a limiting length scale for data from the galaxy surveys. As with the other works we have compared to here, the forecasts for $\sum m_{\nu}$ are very competitive, and are stronger than ours for the DESI surveys with \textit{Planck}. Given the constraining power of CMB measurements in combination with galaxy survey data obtained over a large sky volume, we can still expect it will not be complimented by peculiar velocity measurements, as CMB-S4 will allow for very strong constraints with these other galaxy surveys already. However, CMB-S4, as an early-Universe probe, may allow for interesting comparison between early and late-Universe $\sum m_{\nu}$ constraints, if we compare the constraints to a combination of PV measurements with galaxy redshifts, and weak lensing data (late-Universe probes).

\subsection{Results with \texorpdfstring{$N_{\text{eff}}$}{t} free}
\label{sec:neff}

We move now to considering constraints on an extended cosmological model when the effective number of neutrino species, $N_{\text{eff}}$ is also allowed to vary in addition to the sum of their masses. Table~\ref{results_table_1_nefffree} shows our corresponding forecasts for various surveys combined with our estimate of the Fisher information from the \textit{Planck} 2018 data. Analogous to the earlier Table~\ref{corrected_version_table3}, Table \ref{results_table_2_combined_surveys_nefffree} shows combinations of different surveys with and without the inclusion of our estimate of \textit{Planck} 2018, but now with varying $N_{\text{eff}}$. 

\begin{table}
    \caption{Forecasts for surveys with an estimate of the Fisher information from the \textit{Planck} 2018 results, for $k_{\text{max}} = 0.1h (0.2h) \mathrm{Mpc}^{-1}$, where $N_{\text{eff}}$ is also a free parameter in the forecasting analysis. Unlike in the previous results tables, we don't include information from \textit{Planck} lensing. $\alpha$ gives the assumed fractional uncertainty on our distance indicator for SNe Ia in these forecasts (we have forecasts for two values of $\alpha$ here), the fractional uncertainty on distance indicators for PV surveys other than LSST are as specified in Table \ref{table1_surveydetails}. The second column in this table shows the predicted uncertainty on $\sum m_{\nu}$, $\sigma_{ \sum m_{\nu} }$. The third column shows the relative improvement when adding information from the PV survey (abbreviated red+PV) compared to the corresponding redshift-only survey (red). 
    }
    \centering
    \begin{tabular}{p{3.9cm}|c|c}
    \hline
    \textbf{Surveys} (+Planck 2018) & $\sigma_{\sum m_{\nu}}$ (eV) & 
    $1 - \frac{\sigma_{\text{red+PV}}}{ \sigma_{\text{red}}} \left( \% \right)$ \\ \hline \hline
    DESI BGS & 0.095 (0.086) & -- \\ 
    DESI PVs & 0.105 (0.104) & -- \\ 
    DESI (BGS + PVs) & 0.093 (0.084) & 2.1 (2.3) \\ 
    WALLABY redshifts &  0.104 (0.099) & -- \\ 
    WALLABY PVs &   0.105 (0.104) & -- \\ 
    WALLABY (redshifts and PVs) &  0.091 (0.089) & 12.5 (10.1) \\
    4HS redshifts & 0.098 (0.089) & -- \\ 
    4HS PVs & 0.104 (0.104) & -- \\
    4HS (redshifts and PVs) & 0.095 (0.085) & 3.1 (4.5) \\
    $J < 19$  & 0.093 (0.082) & -- \\
    LSST PVs ($\alpha = 0.05$) & 0.105 (0.104) & -- \\ 
    LSST PVs ($\alpha = 0.1$) & 0.105 (0.105) & -- \\ 
    $J < 19$ + LSST PVs ($\alpha = 0.05$) &  0.092 (0.080) & 1.0 (2.4) \\ 
    $J < 19$ + LSST PVs ($\alpha = 0.1$) & 0.093 (0.082) & 0 (0) \\ 
    DESI Baseline & 0.083 (0.056) & -- \\ 
    \hline 
    \end{tabular}
    \label{results_table_1_nefffree}
\end{table}

\begin{table*}
    \caption{Forecasts for constraints on $\sum m_{\nu}$ for different combinations of datasets, for $k_{\text{max}} = 0.1h (0.2h) \mathrm{Mpc}^{-1}$, where $N_{\text{eff}}$ is also a free parameter in the forecasting analysis.
    Pairs of columns show results with and without the inclusion of information from \textit{Planck} (with no lensing information). $\alpha$ gives the assumed fractional uncertainty on the distances to SNe Ia used for LSST PV forecasts. For other PV surveys, Table \ref{table1_surveydetails} specifies the assumed uncertainty on the distance indicator for PVs. The second and fourth columns show the predicted uncertainty on the sum of neutrino masses, while the third and fifth columns show the relative improvement on the uncertainty for each combination compared to the DESI Baseline survey (DESI BL) alone. 
    }
    \label{results_table_2_combined_surveys_nefffree}
    \centering
    \renewcommand{\arraystretch}{1.3}
    \begin{tabular}{p{8.cm}|c|c|c|c}
    \hline
    \multirow{2}{*}{\textbf{Surveys}} & \multicolumn{2}{c}{\textbf{With \textit{Planck}}} & \multicolumn{2}{c}{\textbf{Without \textit{Planck}}} \\
    & $\sigma_{\sum m_{\nu}}$ (eV) & $1 - \frac{\sigma}{ \sigma_{\text{DESI BL}}} \left(\%\right)$ & $\sigma_{\sum m_{\nu}}$ (eV) & $1 - \frac{\sigma}{ \sigma_{\text{DESI BL}} } \left(\%\right)$ \\ \hline \hline
    DESI Baseline & 0.083 (0.056) & -- & 0.902 (0.397) & -- \\
    DESI Baseline + BGS & 0.081 (0.051) & 2.4 (8.9) & 0.765 (0.341) & 15.2 (14.1) \\
    DESI BL + BGS + DESI PVs & 0.080 (0.051) & 3.6 (8.9)  & 0.345 (0.159)  & 61.8 (59.9) \\
    DESI BL + BGS + $J < 19$ & 0.079 (0.048) & 4.8 (14.3) & 0.665 (0.291) & 26.3 (26.7) \\ 
    DESI BL + BGS + DESI PVs + $J < 19$ & 0.078 (0.048) & 6.0 (14.3) & 0.316 (0.147) & 65.0 (63.0) \\
    DESI BL + BGS + DESI PVs + LSST PVs ($\alpha = 0.05$) & 0.080 (0.051) & 3.6 (8.9) & 0.337 (0.152) & 62.6 (61.7) \\
    \hline 
    \end{tabular}
\end{table*}

\begin{figure*}
    \centering
    \includegraphics[scale=0.62]{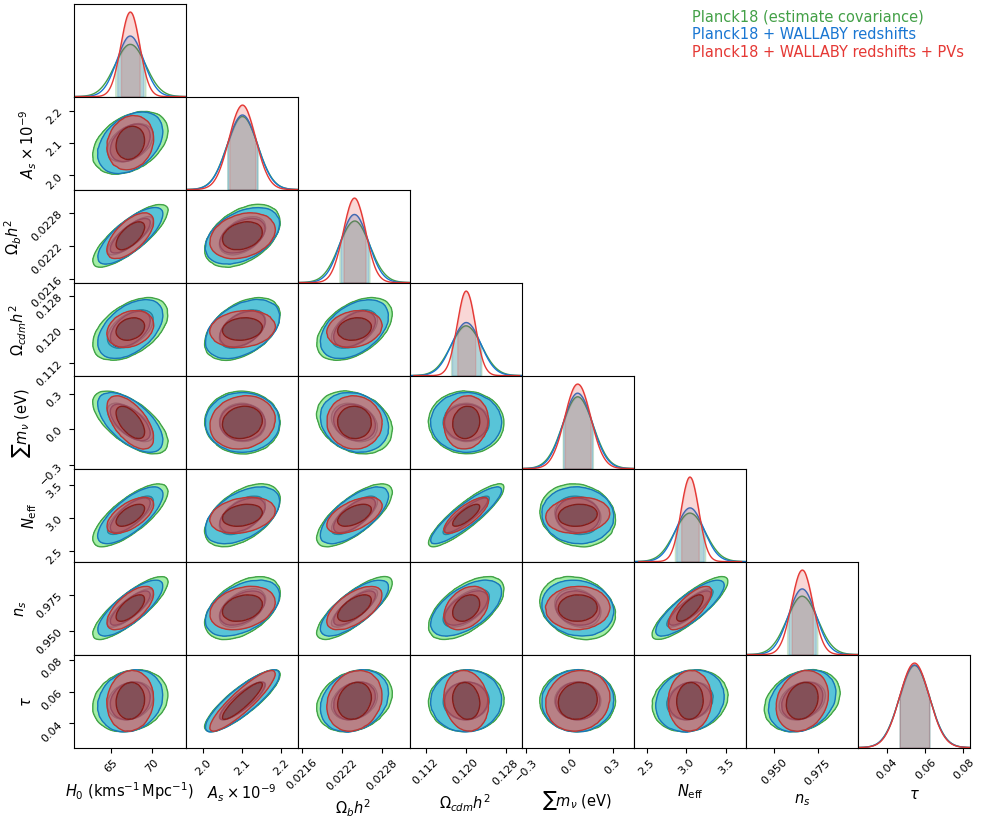}
    \caption{1 and 2-$\sigma$ contours and marginalised likelihoods for cosmological parameters, for our estimate of the Fisher information from the \textit{Planck} 2018 data, when $N_{\text{eff}}$ is free, in green (no lensing information), then the contours when we further add the Fisher information from the WALLABY redshift survey for $k_{\text{max}} = 0.2 h \mathrm{Mpc}^{-1}$ (blue contours), and then with the further addition of the WALLABY peculiar velocity survey information, also for $k_{\text{max}} = 0.2 h \mathrm{Mpc}^{-1}$ (red contours). We also show contours for $\tau$ although $\tau$ is not included as a free parameter in our forecasting analysis, because there is information on $\tau$ from the \textit{Planck} data, which is degenerate with varying $\sum m_{\nu}$ for CMB data. }
    \label{Figure8_contours_nefffree}
\end{figure*}

In Table \ref{results_table_1_nefffree}, there is slightly better improvement on the constraints we forecast from different surveys when information from peculiar velocities is included for most cases (compared to when $N_{\text{eff}}$ is not fixed).In this table, the best forecast overall comes from the DESI Baseline survey regardless of the choice of $k_{\text{max}}$, as was the case for when $N_{\text{eff}}$ was fixed. The inclusion of $N_{\text{eff}}$ as an additional free parameter that is partially degenerate with $\sum m_{\nu}$, unsurprisingly, weakens the constraints in all cases. Figure \ref{Figure8_contours_nefffree}, shows the effects of adding WALLABY peculiar velocities to the WALLABY redshift constraints and \textit{Planck} on all our cosmological free parameters in this analysis, including $N_{\mathrm{eff}}$. 

\begin{figure*}
    \centering
    \includegraphics[scale=0.45]{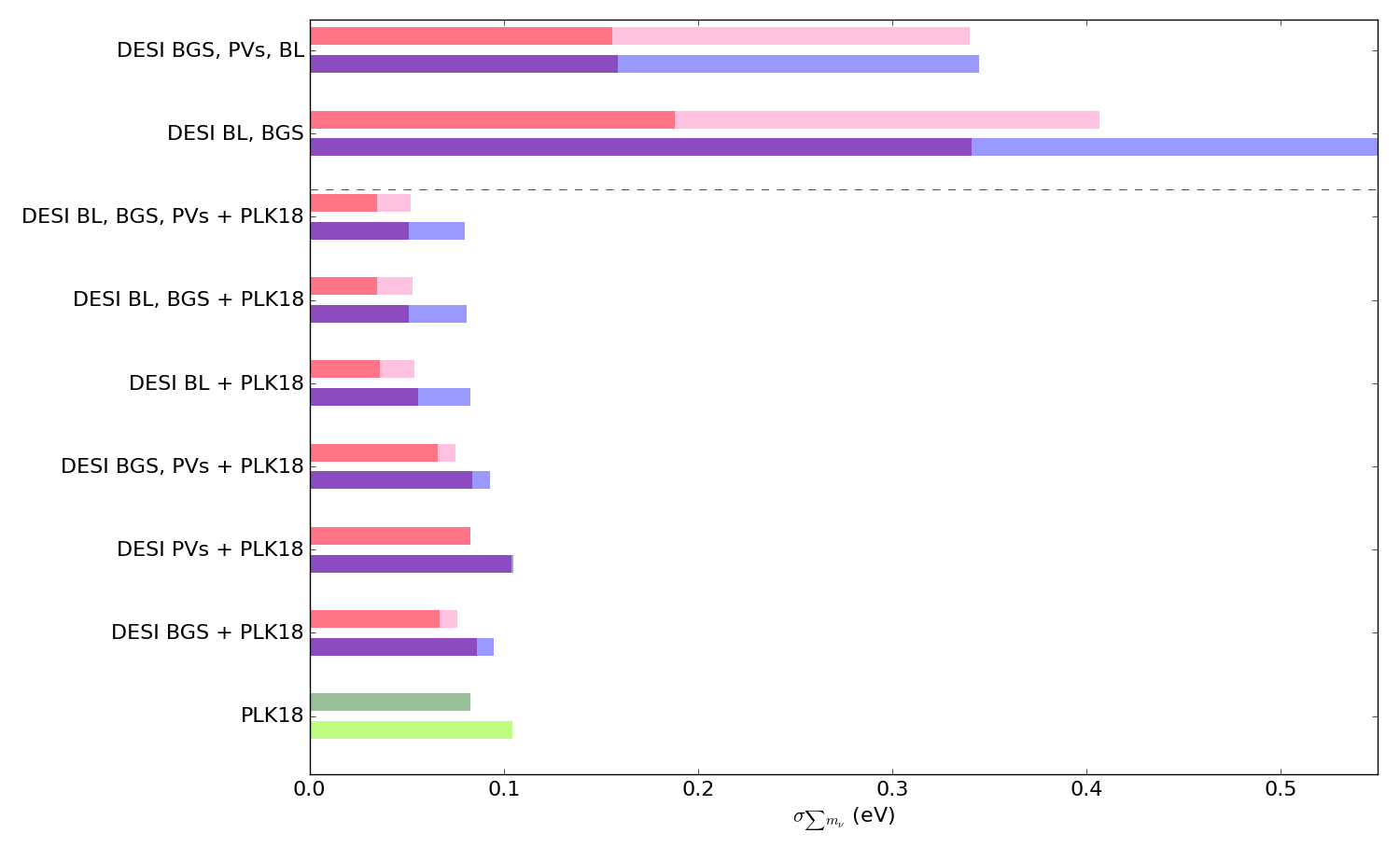}
    \caption{Comparing forecasts from the results of this work when $N_{\text{eff}}$ is a free parameter in the forecasting analysis (blue/purple bars, for $k_{\text{max}} = 0.1h/0.2h \mathrm{Mpc}^{-1}$), to forecasts for the same surveys when $N_{\text{eff}}$ is fixed with the same information from \textit{Planck} (red/pink bars, for $k_{\text{max}} = 0.1h/0.2h \mathrm{Mpc}^{-1}$) to our estimate of the Fisher information in the Planck 2018 results (green when $N_{\text{eff}}$ is free, dark green when $N_{\text{eff}}$ is fixed). Information from the DESI Baseline survey is abbreviated to DESI BL, peculiar velocities to PV, our estimate of the \textit{Planck} 2018 TTTEEE + low$\ell$ + lowE information (no lensing) to PLK18. Forecasts above the dashed line do not include our estimate of the Fisher information from \textit{Planck}. Where the bars go off the plot we have not shown the full range, because it takes away from the small differences that can be seen in the other results, but the reader may assume the values of the uncertainties shown are significantly larger than $0.5$ eV (refer to Table \ref{results_table_2_combined_surveys_nefffree} for numerical values). 
    }
    \label{Figure9_results_barchart_nefffree}
\end{figure*}

Considering Table~\ref{results_table_2_combined_surveys_nefffree}, which is summarised graphically in Figure~\ref{Figure9_results_barchart_nefffree}, there is very small improvement of $\sim 1\%$ when a PV sample is added to the DESI Baseline+BGS samples with \textit{Planck} when $k_{\text{max}} = 0.1h \mathrm{Mpc}^{-1}$, and there is no improvement for this case when $k_{\text{max}} = 0.2h \mathrm{Mpc}^{-1}$. There is better improvement by adding the $J < 19$ sample in either case ($\sim 2-5\%$), and no further improvement by another adding PV sample. When we do not include \textit{Planck}, our results are again similar to the results with $N_{\text{eff}}$ fixed in Table~\ref{corrected_version_table3}. Our results demonstrate that adding information from the DESI PVs to DESI Baseline+BGS, which more than halves the uncertainty on $\sum m_{\nu}$ for either choice of $k_{\text{max}}$, is more effective at improving our forecasts than adding information from the $J < 19$ sample, which has less than half the improvement in each case. These results indicate we are not able to obtain constraints comparable to those of \textit{Planck} with only these survey combinations, but we have showed the addition of the peculiar velocity information allows substantial improvement on $\sum m_{\nu}$ constraints when we vary $N_{\text{eff}}$, compared to adding information from another redshift survey. The addition of other late-Universe probes to these forecasts may potentially allow for even stronger constraints that would be interesting to compare to \textit{Planck}, or other CMB experiments. 

\section{Discussion} \label{sec:discussion}

\subsection{Nuisance parameters}

For the results presented in the previous section, we have assumed to have no knowledge of nuisance parameters by allowing them to vary freely, with no prior information, in each redshift bin. One can reasonably ask whether peculiar velocity information contains useful information on $\sum m_{\nu}$ to combine with redshift survey data, in the scenario we have better knowledge of the galaxy bias $b_g$ or the damping parameter $\sigma_g$ that afflicts redshift surveys? This is because, in such a case, we might expect that having more information about these nuisance parameters would allow for much tighter constraints by redshift surveys on their own and thus reduce the improvement that is gained by adding peculiar velocity information. Therefore we consider the following scenario; if we are very optimistic and assume we had total knowledge of either of these parameters for the DESI BGS survey or the DESI Baseline survey, does the DESI peculiar velocity survey continue to be a worthwhile addition to these datasets for $\sum m_{\nu}$ constraints, or does the improvement from adding this information decrease? 

Table \ref{results:removingnuisanceparams} summarise the forecasts we would obtain for combinations of redshift surveys with and without peculiar velocity data when we assume total knowledge of the galaxy bias for each tracer. We are still able to obtain some improvement on the constraints from the DESI BGS survey with \textit{Planck} when we add DESI peculiar velocities. The peculiar velocity data is contributing information on $\sum m_{\nu}$ that is independent of the fact that peculiar velocity data helps to breaks the degeneracy between $b_g$ and other parameters, when we do not have total information on $b_g$. The result is that we obtain roughly the same percent improvement when adding DESI PVs to the DESI BGS survey with \textit{Planck} of $\sim1-2\%$. 

The improvement when adding DESI PVs to the DESI BGS and Baseline survey (without \textit{Planck}) is now zero, which is weaker than our forecasts when $b_g(z)$ is assumed to be unknown for the tracers in these redshift surveys, where the PV data helps to break the degeneracy between $\Omega_m$ and $f(k)$ with $b_g(z)$ and considerably reduces the uncertainty on $\sum m_{\nu}$.  In this case, the PV data adds no useful information on $\sum m_{\nu}$ due to their low signal-to-noise ratio compared to these redshift surveys when $b_g(z)$ for each of the four redshift tracers is known.

\begin{table}
    \caption{Forecasts for constraints on $\sum m_{\nu}$ for different surveys, for $k_{\text{max}} = 0.2h \mathrm{Mpc}^{-1}$, where we assume to have total knowledge of each the nuisance parameters in our analysis one at a time, by not including it as a free parameter (e.g. remove $b_g(z)$ as a free parameter but continue to vary $\sigma_g(z)$, $\sigma_u(z)$). The third column in this table shows the relative improvement when adding PVs compared to the same datasets without the PV information. 
    } \label{results:removingnuisanceparams}
    \centering
    \begin{tabular}{p{3.9cm}|c|c}
    \hline 
    \textbf{Surveys} & $\sigma_{\sum m_{\nu}}$ (eV) & 
    $1 - \frac{\sigma_{\text{red+PV}}}{ \sigma_{\text{red}} } \left(\%\right)$ \\ \hline \hline 
    $b_g(z)$ removed from analysis \\ 
    DESI BGS + \textit{Planck} & 0.049 & --  \\
    DESI BGS + DESI PVs + \textit{Planck} & 0.048 & 2.0 \\ 
    DESI Baseline + BGS & 0.089 & -- \\ 
    DESI Baseline + BGS + DESI PVs & 0.089 & 0.0 \\  
    $\sigma_g(z)$ removed from analysis \\ 
    DESI BGS + \textit{Planck} & 0.059 & --  \\ 
    DESI BGS + DESI PVs + \textit{Planck} & 0.059 & 0.0  \\
    DESI Baseline + BGS & 0.171 & -- \\ 
    DESI Baseline + BGS + DESI PVs & 0.150 & 12.3 \\ 
    $\sigma_u(z)$ removed from analysis \\
    DESI PVs + \textit{Planck} & 0.077 & --  \\
    DESI BGS + \textit{Planck} & 0.060 & -- \\ 
    DESI BGS + DESI PVs + \textit{Planck} & 0.059 & 1.7 \\ 
    DESI Baseline + DESI BGS &  0.188 & -- \\ 
    DESI Baseline + BGS + DESI PVs & 0.145 & 32.9 \\
    \hline
    \end{tabular}
\end{table}

In our results where we assume full knowledge regarding $\sigma_g(z)$, there is an improvement of $\sim 12\%$ when adding DESI PVs to the DESI Baseline+DESI BGS, although there is still no improvement when adding DESI PVs to DESI BGS+\textit{Planck}. The information from \textit{Planck}+BGS when $\sigma_g(z)$ is known is constraining enough such that the PV data has a signal-to-noise ratio too low to contribute useful information, but this is not the case when \textit{Planck} data is not available. 

Peculiar velocity surveys are also, like redshift surveys, affected by our lack of knowledge about the damping parameter $\sigma_u$ (which is an analogous parameter to $\sigma_g$, see Section \ref{sec:freeparameters} for more detail). We also consider the scenario in Table \ref{results:removingnuisanceparams} in which we have total knowledge of this parameter - how much better is the improvement on $\sum m_{\nu}$ constraints by redshifts surveys, when including peculiar velocity information, if we have total knowledge of $\sigma_u$? The results show there is roughly the same improvement when adding DESI PVs to DESI BGS+\textit{Planck}, but a greater improvement of $\sim 30\%$ in the scenario we consider DESI Baseline+BGS+PVs. 

Given the results in Table \ref{results:removingnuisanceparams}, we can see that having greater knowledge of galaxy bias $b_g$ has the largest affect on any of the constraints on $\sum m_{\nu}$,  when there is considerable information available from redshift tracers, compared to when all nuisance parameters vary freely (see Table \ref{results_table_1_neff_fixed}, \ref{corrected_version_table3}). This is largely due to the fact that galaxy bias is present on all scales and degenerate with $\Omega_{m}$ and the growth rate of structure, both of which are modulated in the presence of massive neutrinos, whereas the damping parameters $\sigma_g$ and $\sigma_u$ only affect the galaxy redshift and galaxy velocity power spectra at non-linear, noisy, scales respectively. 

\subsection{Systematics}

Here we mention potential systematics in redshift and peculiar velocity surveys that we have not accounted for in our analysis, but may have potential to influence the ability of surveys to constrain $\sum m_{\nu}$, or instead bias constraints on $\sum m_{\nu}$ for future surveys. \citet{howlett2017cosmological} analysed the effect of an incorrect velocity zero-point to offset constraints on $f\sigma_8$ (the growth rate of structure $f$ multiplied by the the normalization of the power spectrum), from redshift and peculiar velocity data, and the effect of a scale-dependent velocity bias or galaxy bias. 

\citet{howlett2017cosmological} found that the not accounting for a scale dependent galaxy bias may bias the results for constraints on $f\sigma_8$ by an amount that is less than $0.2\sigma$ for galaxy survey data, although when peculiar velocity data is included, which breaks the degeneracy between cosmological parameters and galaxy bias, it tends to increase the amount of biasing on $f\sigma_8$ such that is may be as large as 1$\sigma$. However, their analysis found that while this systematic may shift the recovered value of $f\sigma_8$ from an analysis, the constraining power of the surveys is typically unaffected (see Table 5 in their work). This systematic is absorbed into the free value of $\sigma_{g}$. Given this analysis, we might expect our forecasts for $\sigma_{\sum m_{\nu}}$, which we have produced assuming a scale independent galaxy bias but while allowing the $\sigma_{g}$ nuisance parameter to vary freely, to also be robust to scale-dependent galaxy bias.

We have assumed for our forecasts here that there is no velocity bias. Including a velocity bias would introduce an extra free parameter to our forecasting analysis, and likely weaken the constraining power of any of our forecasts for velocity surveys alone. However, as we have demonstrated that fixing or varying $\sigma_{u}$, a parameter which affects the power spectrum on similar scales, has little impact on our forecasts for any combinations of redshift and peculiar velocity data when \textit{Planck} data is included, it is reasonable to assume velocity bias would also have a small effect. In the case \textit{Planck} is not included, having more knowledge of $\sigma_u$ actually improved the constraining power from peculiar velocities however. Therefore, considering an extra free parameter to model velocity bias, which affects the power spectrum on similar scales to $\sigma_u$, implies it has the potential to weaken our forecasts in some scenarios. Therefore, we suggest that understanding how the presence of a velocity bias would influence constraints on $\sum m_{\nu}$ would be worth exploring in further work. 

Other systematics may involve an incorrect calibration of the zero-point offset used in the relation used to infer the velocity of each galaxy (i.e. Tully-Fisher or Fundamental Plane relationship). \citet{howlett2017cosmological} showed an offset of the size expected from realistic catalogues had little effect on constraints by galaxy and velocity data for $f\sigma_8$. We hence expect it to have only a small effect on neutrino mass constraints.

To summarise, previous work by \citet{howlett2017cosmological} suggests the value of $\sigma_{\sum m_{\nu}}$ would be largely unaffected by some of the systematics we discuss here, but it might nonetheless be worth investigating this in future work, particularly when one considers a velocity bias and when analysis methods are applied to real velocity data with the aim of constraining the sum of the neutrino masses. 

\section{Conclusions}
\label{sec:conclusions}

To conclude this paper, we address the two questions we asked at the end of Section \ref{intro}. Firstly, does data obtained via measurements of the peculiar motions of galaxies improve constraints on $\sum m_{\nu}$ compared to redshift measurements alone? The results in this work would indicate yes, the peculiar motions of galaxies contain some accessible information that allows us to improve our constraints $\sum m_{\nu}$ that may be obtained from redshift surveys (in addition to information we obtain from measurements of the CMB). We have confirmed this for a number of different configurations for combined redshift and peculiar velocity surveys. The improvement is in some cases, more appreciable when we are limited to $k_{\text{max}} = 0.1h\,\mathrm{Mpc}^{-1}$, although it is more so when $N_{\text{eff}}$ is allowed to be a free parameter, and the improvement is notable when we do not rely on the inclusion of information from \textit{Planck} measurements of the CMB. Allowing $N_{\text{eff}}$ to be a free parameter may be particularly interesting in the context of the Hubble tension (a potential solution to the problem involves allowing for extra species of neutrinos, i.e. `dark radiation', \citep{di2021realm}, and $N_{\text{eff}}$ is particularly degenerate with $\sum m_{\nu}$). Overall, our results show that it is not possible to achieve constraints as strong as those of the CMB measurements by \textit{Planck} (which is able to achieve a 1$\sigma$ constraint of $\sigma_{\sum m_{\nu}} = 0.077$ eV alone using our approximation of the Fisher information) when considering just combinations of upcoming redshift and PV surveys. The best constraint we are able to find for any given combination of redshift or PV surveys we consider is $\sigma_{\sum m_{\nu}} = 0.139$ eV. 

Our second question we aimed to address was, is information from galaxy peculiar motions useful for obtaining better constraints on $\sum m_{\nu}$, such that we should pursue their measurements for this purpose? In the case we add PV information to a combination of \textit{Planck} data with high redshift galaxy surveys rather than low redshift surveys, the effect of adding PVs is negligible. While we are still able to show substantial improvement to neutrino mass constraints when we do not include information from \textit{Planck} to redshift and PV surveys, and we show that adding PV surveys in this scenario allows for better improvement to neutrino mass constraints than including information from more redshift surveys, current constraints from \textit{Planck} are considerably stronger. To achieve constraints comparable to those that are possible from \textit{Planck}, future CMB experiments or these in combination with upcoming redshift surveys, will require the addition of further data from other low redshift probes. This might allow for an independent comparison of constraints on $\sum m_{\nu}$ between early-Universe probes and late-Universe probes that would be interesting to consider in future work, and would be relevant for current cosmological parameter tensions.

\section*{Acknowledgements}
The authors thank Khaled Said for useful discussions. We also thank Christoph Saulder and Kelly Douglass for input on producing the target densities for the DESI peculiar velocity survey, and Edward Taylor and Michelle Cluver for doing the same for the 4MOST Hemisphere Survey. CH and TMD acknowledge support from the Australian Government through the Australian Research Council’s Laureate Fellowship funding scheme (project FL180100168). This research has made use of NASA’s Astrophysics Data System Bibliographic Services and the astro-ph pre-print archive at \href{https://arxiv.org/}{https://arxiv.org/}, the matplotlib plotting library \citep{Hunter:2007}, and the chainconsumer package \citep{Hinton2016}.

\section*{Data Availability}
The code and data used for all forecasts in this work is available at \href{https://github.com/abbew25/PV_galaxy_survey_forecasts_neutrinos}{https://github.com/abbew25/PV\_galaxy\_survey\_forecasts\_neutrinos}.



\bibliographystyle{mnras}
\bibliography{neutrinos_paper} 

\begin{thebibliography}{}
\makeatletter
\relax
\def\mn@urlcharsother{\let\do\@makeother \do\$\do\&\do\#\do\^\do\_\do\%\do\~}
\def\mn@doi{\begingroup\mn@urlcharsother \@ifnextchar [ {\mn@doi@}
  {\mn@doi@[]}}
\def\mn@doi@[#1]#2{\def\@tempa{#1}\ifx\@tempa\@empty \href
  {http://dx.doi.org/#2} {doi:#2}\else \href {http://dx.doi.org/#2} {#1}\fi
  \endgroup}
\def\mn@eprint#1#2{\mn@eprint@#1:#2::\@nil}
\def\mn@eprint@arXiv#1{\href {http://arxiv.org/abs/#1} {{\tt arXiv:#1}}}
\def\mn@eprint@dblp#1{\href {http://dblp.uni-trier.de/rec/bibtex/#1.xml}
  {dblp:#1}}
\def\mn@eprint@#1:#2:#3:#4\@nil{\def\@tempa {#1}\def\@tempb {#2}\def\@tempc
  {#3}\ifx \@tempc \@empty \let \@tempc \@tempb \let \@tempb \@tempa \fi \ifx
  \@tempb \@empty \def\@tempb {arXiv}\fi \@ifundefined
  {mn@eprint@\@tempb}{\@tempb:\@tempc}{\expandafter \expandafter \csname
  mn@eprint@\@tempb\endcsname \expandafter{\@tempc}}}

\bibitem[\protect\citeauthoryear{Abazajian et~al.,}{Abazajian
  et~al.}{2019}]{abazajian2019cmb}
Abazajian K.,  et~al., 2019, arXiv preprint arXiv:1907.04473

\bibitem[\protect\citeauthoryear{Abe et~al.,}{Abe
  et~al.}{2018}]{abe2018atmospheric}
Abe K.,  et~al., 2018, Physical Review D, 97, 072001

\bibitem[\protect\citeauthoryear{Acero et~al.,}{Acero
  et~al.}{2019}]{acero2019first}
Acero M.,  et~al., 2019, Physical review letters, 123, 151803

\bibitem[\protect\citeauthoryear{Adey et~al.,}{Adey
  et~al.}{2018}]{adey2018measurement}
Adey D.,  et~al., 2018, Physical review letters, 121, 241805

\bibitem[\protect\citeauthoryear{Aghamousa et~al.,}{Aghamousa
  et~al.}{2016}]{aghamousa2016desi}
Aghamousa A.,  et~al., 2016, arXiv preprint arXiv:1611.00036

\bibitem[\protect\citeauthoryear{Aghanim et~al.,}{Aghanim
  et~al.}{2020}]{aghanim2020planck}
Aghanim N.,  et~al., 2020, Astronomy \& Astrophysics, 641, A6

\bibitem[\protect\citeauthoryear{Agrawal, Okumura  \& Futamase}{Agrawal
  et~al.}{2019}]{agrawal2019constraining}
Agrawal A.,  Okumura T.,   Futamase T.,  2019, Physical Review D, 100, 063534

\bibitem[\protect\citeauthoryear{Alam et~al.,}{Alam
  et~al.}{2021}]{alam2021completed}
Alam S.,  et~al., 2021, Physical Review D, 103, 083533

\bibitem[\protect\citeauthoryear{Alcock \& Paczy{\'n}ski}{Alcock \&
  Paczy{\'n}ski}{1979}]{alcock1979evolution}
Alcock C.,  Paczy{\'n}ski B.,  1979, Nature, 281, 358

\bibitem[\protect\citeauthoryear{Allison, Caucal, Calabrese, Dunkley  \&
  Louis}{Allison et~al.}{2015}]{allison2015towards}
Allison R.,  Caucal P.,  Calabrese E.,  Dunkley J.,   Louis T.,  2015, Physical
  Review D, 92, 123535

\bibitem[\protect\citeauthoryear{Amendola \& Quartin}{Amendola \&
  Quartin}{2021}]{amendola2021measuring}
Amendola L.,  Quartin M.,  2021, Monthly Notices of the Royal Astronomical
  Society, 504, 3884

\bibitem[\protect\citeauthoryear{{Archidiacono}, {Hannestad}  \&
  {Lesgourgues}}{{Archidiacono} et~al.}{2020}]{Archidiacono2020}
{Archidiacono} M.,  {Hannestad} S.,   {Lesgourgues} J.,  2020, \mn@doi [\jcap]
  {10.1088/1475-7516/2020/09/021}, \href
  {https://ui.adsabs.harvard.edu/abs/2020JCAP...09..021A} {2020, 021}

\bibitem[\protect\citeauthoryear{Bernal, Smith, Boddy  \& Kamionkowski}{Bernal
  et~al.}{2020}]{bernal2020robustness}
Bernal J.~L.,  Smith T.~L.,  Boddy K.~K.,   Kamionkowski M.,  2020, Physical
  Review D, 102, 123515

\bibitem[\protect\citeauthoryear{Blas, Lesgourgues  \& Tram}{Blas
  et~al.}{2011}]{blas2011cosmic}
Blas D.,  Lesgourgues J.,   Tram T.,  2011, Journal of Cosmology and
  Astroparticle Physics, 2011, 034

\bibitem[\protect\citeauthoryear{{Burkey} \& {Taylor}}{{Burkey} \&
  {Taylor}}{2004}]{Burkey2004}
{Burkey} D.,  {Taylor} A.~N.,  2004, \mn@doi [\mnras]
  {10.1111/j.1365-2966.2004.07192.x}, \href
  {https://ui.adsabs.harvard.edu/abs/2004MNRAS.347..255B} {347, 255}

\bibitem[\protect\citeauthoryear{Castro, Quartin  \& Benitez-Herrera}{Castro
  et~al.}{2016}]{castro2016turning}
Castro T.,  Quartin M.,   Benitez-Herrera S.,  2016, Physics of the dark
  universe, 13, 66

\bibitem[\protect\citeauthoryear{Colas, d'Amico, Senatore, Zhang  \&
  Beutler}{Colas et~al.}{2020}]{colas2020efficient}
Colas T.,  d'Amico G.,  Senatore L.,  Zhang P.,   Beutler F.,  2020, Journal of
  Cosmology and Astroparticle Physics, 2020, 001

\bibitem[\protect\citeauthoryear{{Desjacques} \& {Sheth}}{{Desjacques} \&
  {Sheth}}{2010}]{Desjacques2010}
{Desjacques} V.,  {Sheth} R.~K.,  2010, \mn@doi [\prd]
  {10.1103/PhysRevD.81.023526}, \href
  {https://ui.adsabs.harvard.edu/abs/2010PhRvD..81b3526D} {81, 023526}

\bibitem[\protect\citeauthoryear{{Dey} et~al.,}{{Dey} et~al.}{2019}]{Dey2019}
{Dey} A.,  et~al., 2019, \mn@doi [\aj] {10.3847/1538-3881/ab089d}, \href
  {https://ui.adsabs.harvard.edu/abs/2019AJ....157..168D} {157, 168}

\bibitem[\protect\citeauthoryear{Di~Valentino et~al.,}{Di~Valentino
  et~al.}{2021}]{di2021realm}
Di~Valentino E.,  et~al., 2021, arXiv preprint arXiv:2103.01183

\bibitem[\protect\citeauthoryear{Djorgovski \& Davis}{Djorgovski \&
  Davis}{1987}]{djorgovski1987fundamental}
Djorgovski S.,  Davis M.,  1987, Astrophysical Journal, 313, 59

\bibitem[\protect\citeauthoryear{{Driver} et~al.,}{{Driver}
  et~al.}{2019}]{Driver2019}
{Driver} S.~P.,  et~al., 2019, \mn@doi [The Messenger]
  {10.18727/0722-6691/5126}, \href
  {https://ui.adsabs.harvard.edu/abs/2019Msngr.175...46D} {175, 46}

\bibitem[\protect\citeauthoryear{Duffy, Meyer, Staveley-Smith, Bernyk, Croton,
  Koribalski, Gerstmann  \& Westerlund}{Duffy
  et~al.}{2012}]{duffy2012predictions}
Duffy A.~R.,  Meyer M.~J.,  Staveley-Smith L.,  Bernyk M.,  Croton D.~J.,
  Koribalski B.~S.,  Gerstmann D.,   Westerlund S.,  2012, Monthly Notices of
  the Royal Astronomical Society, 426, 3385

\bibitem[\protect\citeauthoryear{Esteban, Gonz{\'a}lez-Garc{\'\i}a, Maltoni,
  Schwetz  \& Zhou}{Esteban et~al.}{2020}]{esteban2020fate}
Esteban I.,  Gonz{\'a}lez-Garc{\'\i}a M.~C.,  Maltoni M.,  Schwetz T.,   Zhou
  A.,  2020, Journal of High Energy Physics, 2020, 1

\bibitem[\protect\citeauthoryear{Fakhouri et~al.,}{Fakhouri
  et~al.}{2015}]{fakhouri2015improving}
Fakhouri H.,  et~al., 2015, The Astrophysical Journal, 815, 58

\bibitem[\protect\citeauthoryear{{Finoguenov} et~al.,}{{Finoguenov}
  et~al.}{2019}]{Finoguenov2019}
{Finoguenov} A.,  et~al., 2019, \mn@doi [The Messenger]
  {10.18727/0722-6691/5124}, \href
  {https://ui.adsabs.harvard.edu/abs/2019Msngr.175...39F} {175, 39}

\bibitem[\protect\citeauthoryear{Font-Ribera, McDonald, Mostek, Reid, Seo  \&
  Slosar}{Font-Ribera et~al.}{2014}]{font2014desi}
Font-Ribera A.,  McDonald P.,  Mostek N.,  Reid B.~A.,  Seo H.-J.,   Slosar A.,
   2014, Journal of Cosmology and Astroparticle Physics, 2014, 023

\bibitem[\protect\citeauthoryear{Fukuda et~al.,}{Fukuda
  et~al.}{1998}]{fukuda1998evidence}
Fukuda Y.,  et~al., 1998, Physical Review Letters, 81, 1562

\bibitem[\protect\citeauthoryear{{GAMBIT Cosmology Workgroup} et~al.,}{{GAMBIT
  Cosmology Workgroup} et~al.}{2020}]{Stoecker2021}
{GAMBIT Cosmology Workgroup} T.,  et~al., 2020, arXiv e-prints, \href
  {https://ui.adsabs.harvard.edu/abs/2020arXiv200903287G} {p. arXiv:2009.03287}

\bibitem[\protect\citeauthoryear{Gerbino \& Lattanzi}{Gerbino \&
  Lattanzi}{2018}]{gerbino2018status}
Gerbino M.,  Lattanzi M.,  2018, Frontiers in Physics, 5, 70

\bibitem[\protect\citeauthoryear{{Hinton}}{{Hinton}}{2016}]{Hinton2016}
{Hinton} S.~R.,  2016, \mn@doi [The Journal of Open Source Software]
  {10.21105/joss.00045}, \href
  {http://adsabs.harvard.edu/abs/2016JOSS....1...45H} {1, 00045}

\bibitem[\protect\citeauthoryear{{Hong} et~al.,}{{Hong}
  et~al.}{2019}]{Hong2019}
{Hong} T.,  et~al., 2019, \mn@doi [\mnras] {10.1093/mnras/stz1413}, \href
  {https://ui.adsabs.harvard.edu/abs/2019MNRAS.487.2061H} {487, 2061}

\bibitem[\protect\citeauthoryear{Howlett}{Howlett}{2019}]{howlett2019redshift}
Howlett C.,  2019, Monthly Notices of the Royal Astronomical Society, 487, 5209

\bibitem[\protect\citeauthoryear{Howlett, Staveley-Smith  \& Blake}{Howlett
  et~al.}{2017a}]{howlett2017cosmological}
Howlett C.,  Staveley-Smith L.,   Blake C.,  2017a, Monthly Notices of the
  Royal Astronomical Society, 464, 2517

\bibitem[\protect\citeauthoryear{{Howlett} et~al.,}{{Howlett}
  et~al.}{2017b}]{Howlett2017c}
{Howlett} C.,  et~al., 2017b, \mn@doi [\mnras] {10.1093/mnras/stx1521}, \href
  {https://ui.adsabs.harvard.edu/abs/2017MNRAS.471.3135H} {471, 3135}

\bibitem[\protect\citeauthoryear{Howlett, Robotham, Lagos  \& Kim}{Howlett
  et~al.}{2017c}]{howlett2017measuring}
Howlett C.,  Robotham A.~S.,  Lagos C.~D.,   Kim A.~G.,  2017c, The
  Astrophysical Journal, 847, 128

\bibitem[\protect\citeauthoryear{Hunter}{Hunter}{2007}]{Hunter:2007}
Hunter J.~D.,  2007, \mn@doi [Computing in Science \& Engineering]
  {10.1109/MCSE.2007.55}, 9, 90

\bibitem[\protect\citeauthoryear{Ikeda, Collaboration  et~al.}{Ikeda
  et~al.}{2008}]{ikeda2008solar}
Ikeda M.,  Collaboration S.-K.,   et~al., 2008, in Journal of Physics:
  Conference Series. p. 042009

\bibitem[\protect\citeauthoryear{Ivezi{\'c} et~al.,}{Ivezi{\'c}
  et~al.}{2019}]{ivezic2019lsst}
Ivezi{\'c} {\v{Z}}.,  et~al., 2019, The Astrophysical Journal, 873, 111

\bibitem[\protect\citeauthoryear{Kaiser}{Kaiser}{1987}]{kaiser1987clustering}
Kaiser N.,  1987, Monthly Notices of the Royal Astronomical Society, 227, 1

\bibitem[\protect\citeauthoryear{Koda et~al.,}{Koda
  et~al.}{2014}]{koda2014peculiar}
Koda J.,  et~al., 2014, Monthly Notices of the Royal Astronomical Society, 445,
  4267

\bibitem[\protect\citeauthoryear{Koribalski et~al.,}{Koribalski
  et~al.}{2020}]{koribalski2020wallaby}
Koribalski B.~S.,  et~al., 2020, Astrophysics and Space Science, 365, 1

\bibitem[\protect\citeauthoryear{Laureijs et~al.,}{Laureijs
  et~al.}{2012}]{laureijs2012euclid}
Laureijs R.,  et~al., 2012, in Space Telescopes and Instrumentation 2012:
  Optical, Infrared, and Millimeter Wave. p. 84420T

\bibitem[\protect\citeauthoryear{Lesgourgues \& Pastor}{Lesgourgues \&
  Pastor}{2012}]{lesgourgues2012neutrino}
Lesgourgues J.,  Pastor S.,  2012, Advances in High Energy Physics, 2012

\bibitem[\protect\citeauthoryear{Lesgourgues \& Pastor}{Lesgourgues \&
  Pastor}{2014}]{lesgourgues2014neutrino}
Lesgourgues J.,  Pastor S.,  2014, New Journal of Physics, 16, 065002

\bibitem[\protect\citeauthoryear{{Magoulas} et~al.,}{{Magoulas}
  et~al.}{2012}]{Magoulas2012}
{Magoulas} C.,  et~al., 2012, \mn@doi [\mnras]
  {10.1111/j.1365-2966.2012.21421.x}, \href
  {https://ui.adsabs.harvard.edu/abs/2012MNRAS.427..245M} {427, 245}

\bibitem[\protect\citeauthoryear{Marshall et~al.,}{Marshall
  et~al.}{2017}]{marshall2017science}
Marshall P.,  et~al., 2017, arXiv preprint arXiv:1708.04058

\bibitem[\protect\citeauthoryear{{McDonald} \& {Seljak}}{{McDonald} \&
  {Seljak}}{2009}]{Mcdonald2009}
{McDonald} P.,  {Seljak} U.,  2009, \mn@doi [\jcap]
  {10.1088/1475-7516/2009/10/007}, \href
  {https://ui.adsabs.harvard.edu/abs/2009JCAP...10..007M} {2009, 007}

\bibitem[\protect\citeauthoryear{{Merloni} et~al.,}{{Merloni}
  et~al.}{2019}]{Merloni2019}
{Merloni} A.,  et~al., 2019, \mn@doi [The Messenger] {10.18727/0722-6691/5125},
  \href {https://ui.adsabs.harvard.edu/abs/2019Msngr.175...42M} {175, 42}

\bibitem[\protect\citeauthoryear{Palanque-Delabrouille
  et~al.,}{Palanque-Delabrouille et~al.}{2015}]{palanque2015neutrino}
Palanque-Delabrouille N.,  et~al., 2015, Journal of Cosmology and Astroparticle
  Physics, 2015, 011

\bibitem[\protect\citeauthoryear{{Peebles}}{{Peebles}}{1980}]{Peebles1980}
{Peebles} P.~J.~E.,  1980, {The large-scale structure of the universe}.
Princeton University Press

\bibitem[\protect\citeauthoryear{Phillips}{Phillips}{1993}]{phillips1993absolute}
Phillips M.~M.,  1993, The Astrophysical Journal, 413, L105

\bibitem[\protect\citeauthoryear{Quartin, Amendola  \& Moraes}{Quartin
  et~al.}{2021}]{quartin20216x2pt}
Quartin M.,  Amendola L.,   Moraes B.,  2021, arXiv preprint arXiv:2111.05185

\bibitem[\protect\citeauthoryear{Rest et~al.,}{Rest
  et~al.}{2014}]{rest2014cosmological}
Rest A.,  et~al., 2014, The Astrophysical Journal, 795, 44

\bibitem[\protect\citeauthoryear{{Richard} et~al.,}{{Richard}
  et~al.}{2019}]{Richard2019}
{Richard} J.,  et~al., 2019, \mn@doi [The Messenger] {10.18727/0722-6691/5127},
  \href {https://ui.adsabs.harvard.edu/abs/2019Msngr.175...50R} {175, 50}

\bibitem[\protect\citeauthoryear{Riemer-S{\o}rensen, Parkinson  \&
  Davis}{Riemer-S{\o}rensen et~al.}{2013}]{riemer2013half}
Riemer-S{\o}rensen S.,  Parkinson D.,   Davis T.~M.,  2013, Publications of the
  Astronomical Society of Australia, 30

\bibitem[\protect\citeauthoryear{{Ruiz-Macias} et~al.,}{{Ruiz-Macias}
  et~al.}{2020}]{Ruiz-Macias2020}
{Ruiz-Macias} O.,  et~al., 2020, \mn@doi [Research Notes of the American
  Astronomical Society] {10.3847/2515-5172/abc25a}, \href
  {https://ui.adsabs.harvard.edu/abs/2020RNAAS...4..187R} {4, 187}

\bibitem[\protect\citeauthoryear{Sartoris et~al.,}{Sartoris
  et~al.}{2016}]{sartoris2016next}
Sartoris B.,  et~al., 2016, Monthly Notices of the Royal Astronomical Society,
  459, 1764

\bibitem[\protect\citeauthoryear{Sprenger, Archidiacono, Brinckmann, Clesse  \&
  Lesgourgues}{Sprenger et~al.}{2019}]{sprenger2019cosmology}
Sprenger T.,  Archidiacono M.,  Brinckmann T.,  Clesse S.,   Lesgourgues J.,
  2019, Journal of Cosmology and Astroparticle Physics, 2019, 047

\bibitem[\protect\citeauthoryear{Stevens, Croton, Mutch  \& Sinha}{Stevens
  et~al.}{2017}]{stevens2017dark}
Stevens A.~R.,  Croton D.~J.,  Mutch S.~J.,   Sinha M.,  2017, Astrophysics
  Source Code Library, pp ascl--1706

\bibitem[\protect\citeauthoryear{Taylor}{Taylor}{2020}]{taylor2020taipan}
Taylor E.,  2020, The Build-Up of Galaxies through Multiple Tracers and
  Facilities, p.~75

\bibitem[\protect\citeauthoryear{Tully \& Fisher}{Tully \&
  Fisher}{1977}]{tully1977new}
Tully R.~B.,  Fisher J.~R.,  1977, Astronomy and Astrophysics, 54, 661

\bibitem[\protect\citeauthoryear{Vagnozzi, Brinckmann, Archidiacono, Freese,
  Gerbino, Lesgourgues  \& Sprenger}{Vagnozzi et~al.}{2018}]{vagnozzi2018bias}
Vagnozzi S.,  Brinckmann T.,  Archidiacono M.,  Freese K.,  Gerbino M.,
  Lesgourgues J.,   Sprenger T.,  2018, Journal of Cosmology and Astroparticle
  Physics, 2018, 001

\bibitem[\protect\citeauthoryear{Xu, DePorzio, Mu{\~n}oz  \& Dvorkin}{Xu
  et~al.}{2021}]{xu2021accurately}
Xu W.~L.,  DePorzio N.,  Mu{\~n}oz J.~B.,   Dvorkin C.,  2021, Physical Review
  D, 103, 023503

\bibitem[\protect\citeauthoryear{Zheng, Zhang  \& Jing}{Zheng
  et~al.}{2015}]{zheng2015determination}
Zheng Y.,  Zhang P.,   Jing Y.,  2015, Physical Review D, 91, 123512

\bibitem[\protect\citeauthoryear{{de Jong} et~al.,}{{de Jong}
  et~al.}{2019}]{deJong2019}
{de Jong} R.~S.,  et~al., 2019, \mn@doi [The Messenger]
  {10.18727/0722-6691/5117}, \href
  {https://ui.adsabs.harvard.edu/abs/2019Msngr.175....3D} {175, 3}

\makeatother
\end{thebibliography}



\appendix

\section{Numerical derivatives step sizes}\label{ndss}
\begin{figure*}
    \centering
    \includegraphics[scale=0.5]{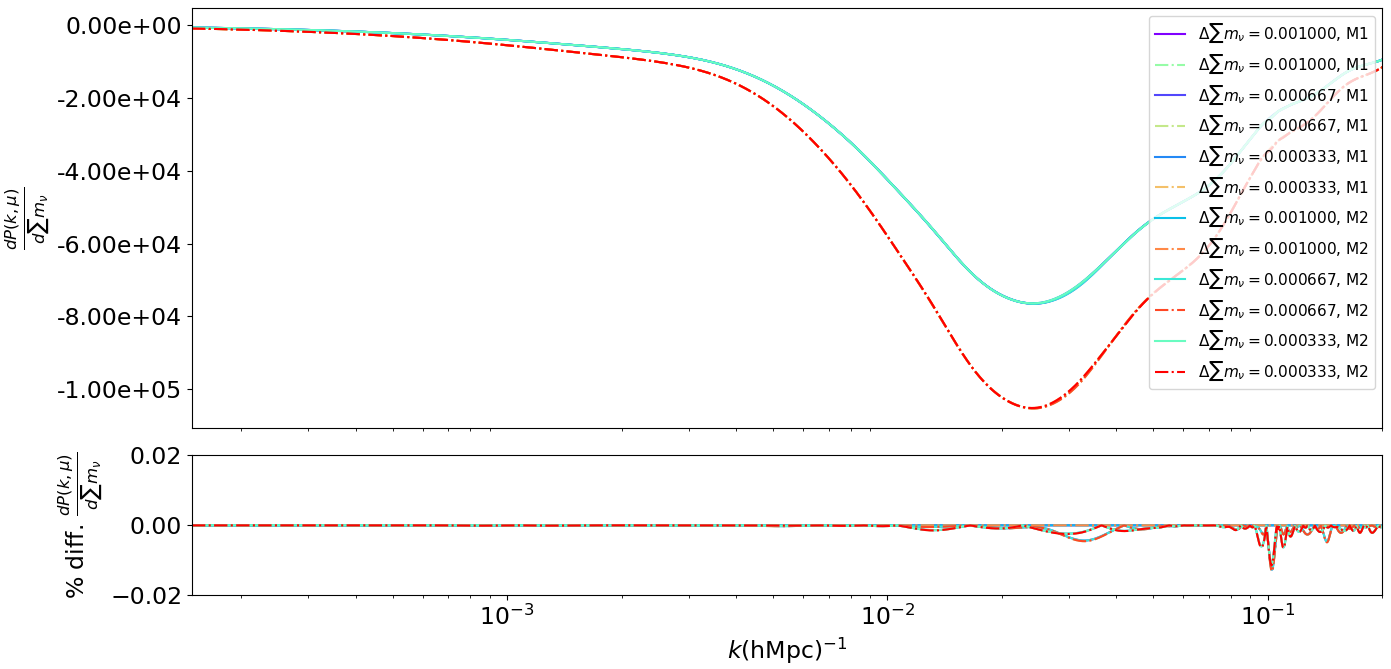}
    \caption{Numerically computed derivatives of the redshift space power spectrum of galaxies with respect to $\sum m_{\nu}$ for different step sizes $\Delta x$, and two different values of $\mu$ at z = 0. The derivatives shown have been computed using two different methods for comparison (M1 = Method 1, M2 = Methods 2), M1 being the case where we finite difference the full galaxy redshift-space power spectra, M2 being the case where we use a semi-analytical approach that involves finite differences of the real space matter power spectrum and growth rate of structure before using the chain rule to compute the galaxy power spectrum derivative. The dashed lines correspond to $\mu$ = 0.7 and the solid lines to $\mu$ = 0.3 for the power
spectra in redshift space. While the top panel shows $\frac{dP}{d\sum m_{\nu}}$, the bottom panel shows the relative difference between the power spectra computed with different step sizes and methods for each value of $\mu$.}
    \label{Figure10_derivative_mnu}
\end{figure*}

In order to confirm the accuracy of the numerical derivatives used in our forecasts, we tested a variety of finite-difference step sizes. Table \ref{table_appendix_derivative_step_sizes} lists the step sizes and central values used in our analysis. For each of the step sizes chosen in these ranges, the computed derivatives of the power spectra were the same, to within 5\%, for all the redshifts and $k$-modes used in this work. Figure \ref{Figure10_derivative_mnu} shows a numerical derivative of the redshift space power spectrum with respect to $\sum m_{\nu}$ about the central value and for the step sizes shown in Table \ref{table_appendix_derivative_step_sizes}. By eye the derivatives are indistinguishable for different step sizes and our two methods of computing them, and the bottom panel shows that the relative differences between the derivatives for different step sizes and methods are less than 2\%. 
\begin{table}
\caption{Tested ranges of step sizes for numerical derivatives of the power spectra, with respect to cosmological parameters, that were found to be robust.}
\centering 
        \begin{tabular}{|c|c|c|} 
            \hline
            $\mathbf{x}_i$ & Valid $\mathbf{\Delta x}_i$ & Central value $\mathbf{x}_i$  \\ \hline \hline
            $H_0$ & 0.018-2.7 (km $\text{s}^{-1} \text{Mpc}^{-1}$) & 67.32 (km $\text{s}^{-1} \text{Mpc}^{-1}$) \\ 
            $A_s$ & $\left(0.0096-0.01\right) \times 10^{-10}$ & 2.10058 $\times 10^{-9}$ \\ 
            $\sum m_{\nu}$ & 0.00033-0.005 (eV) & 0.058 (eV) \\ 
            $\Omega_{\text{b}}h^2$ & 0.00002-0.00075 & 0.022383 \\
            $\Omega_{\text{cdm}}h^2$ & 0.00005-0.006 & 0.12011 \\ 
            $N_{\text{eff}}$ & 0.01-0.075 & 3.046 \\ 
            $ n_s $ & 0.0033-0.01 & 0.99605 \\
            \hline 
        \end{tabular} 
        \label{table_appendix_derivative_step_sizes}
\end{table}

\section{Equations for the Alcock-Paczynski effect}\label{APeq}

The true value of the distance between objects in the line-of-sight and perpendicular direction, denoted $r_{\parallel}^{t}$ and $r_{\perp}^{t}$ in real space respectively, are related to the distance that is computed from a fiducial cosmological model, by the distortion parameters $q_{\parallel}$ and $q_{\perp}$; $r_{\parallel}^{t} = q_{\parallel} r_{\parallel}^{\text{Fid}}$, $r_{\perp}^{t} = q_{\perp} r_{\perp}^{\text{Fid}}$ \citep{bernal2020robustness}. The distortion parameters are defined as \citep{bernal2020robustness}
\begin{align}
    q_{\parallel}(z) & = \frac{D_A(z)H_0}{D_A^{\text{Fid}}(z)H^{\text{Fid}}_0},  \nonumber \\
    q_{\perp}(z) & = \frac{H^{\text{Fid}}(z) H_0}{H(z) H^{\text{Fid}}_0},
\end{align}
where $D_A(z)$ is the comoving angular diameter distance to an object at a redshift $z$. Accounting for the AP effect, an expansion of the power spectrum into Legendre multipoles $\mathcal{P}_l$ can be written as \citep{colas2020efficient},
\begin{align}
    P_l(k) &= \frac{(2l + 1)}{2q_{\parallel}q_{\perp}^2} \int_{-1}^{1} d\mu P\left(k(k^{\text{Fid}}, \mu^{\text{Fid}}, q_{\parallel}, q_{\perp}), \mu(\mu^{\text{Fid}}, q_{\parallel}, q_{\perp})\right) \nonumber \\ & \times \mathcal{P}_l(\mu^{\text{Fid}}).
\end{align}
Overall, the power spectra acquire a factor of $\frac{1}{q_{\parallel} q^2_{\perp}}$ due to the change in comoving volume between the true and fiducial cosmologies, and the value of the power spectrum evaluated at some $\mu$ and $k$ becomes a function of $k^{\text{Fid}}$ and $\mu^{\text{Fid}}$. The wavenumber $k$ and cosine of the angle $\mu$ for a pair of objects that are actually detected in survey data are related to $k^{\text{Fid}}$ and $\mu^{\text{Fid}}$ (the $k$ and $\mu$ the observer believes is correct based on their assumed cosmological model) by the following equations \citep{colas2020efficient},
\begin{equation}
    k = \frac{k^{\text{Fid}}}{q_{\perp}} \left( 1 + (\mu^{\text{Fid}})^2\left( \frac{1}{F^2} - 1 \right) \right)^{1/2},
\end{equation}
\begin{equation}
    \mu = \frac{\mu^{\text{Fid}}}{F} \left( 1 + (\mu^{\text{Fid}})^2\left( \frac{1}{F^2} - 1 \right) \right)^{-1/2},
\end{equation}
where $F = \frac{q_{\parallel}}{q_{\perp}}$.


\bsp	
\label{lastpage}
\end{document}